\documentclass[12pt,a4paper]{article}
\usepackage{amsmath,amssymb,epsfig}
\newcommand{\beq}{\begin{eqnarray}}
\newcommand{\eeq}{\end{eqnarray}}

\newcommand{\be}{\begin{equation}}
\newcommand{\ee}{\end{equation}}

\newcommand{\bm}{\begin{multline}}
\newcommand{\fm}{\end{multline}}

\begin{document}
\setlength{\unitlength}{.8mm}

\begin{titlepage} 
%\begin{flushright}
%version1\\
%\today
%\end{flushright}
\vspace*{0.5cm}
\begin{center}
{\Large\bf Discrete Hirota dynamics for AdS/CFT}
\end{center}
\vspace{1.5cm}
\begin{center}
{\large \'Arp\'ad Heged\H us}
\end{center}
\bigskip

\vspace{0.1cm}

\begin{center}
Research Institute for Particle and Nuclear Physics,\\
Hungarian Academy of Sciences,\\
H-1525 Budapest 114, P.O.B. 49, Hungary\\ 
\end{center}
\vspace{2.5cm}
\begin{abstract}

Recently a set of functional equations defining the anomalous dimensions of arbitrary local
single trace operators in planar ${\cal N}=4$ supersymmetric Yang-Mills theory 
has been conjectured. These functional
equations take the form of a Y-system defined on a special shaped domain. This Y-system can be
equivalently reformulated as a T-system defined on a "T-shaped fat hook". The elements of the
T-system satisfy discrete Hirota equations. In the present paper the discrete Hirota 
equations for AdS/CFT are solved by means of a chain of B\"acklund transformations and 
as a result TT-, TQ-, and QQ-relations are obtained for AdS/CFT.
  
%${\cal Q},{\cal \tilde{Q}}_k^{l,m}$

\end{abstract}

\end{titlepage}

\section{Introduction}

The AdS/CFT correspondence \cite{1} states the equivalence of ${\cal N}=4$ supersymmetric Yang-Mills (SYM) %%@
theory
with superstrings on $AdS_5 \times S^5$. Since the correspondence is a strong-weak duality it is very difficult
to test and prove it. The breakthrough in this respect was the discovery of integrability on both sides of the
duality \cite{2}-\cite{8}. On the string theory side it means that the light-cone quantized worldsheet sigma 
model is an
integrable quantum field theory, while on the gauge theory side integrability manifests itself in the
 appearance of spin chains.

To test and prove AdS/CFT one has to compute
 anomalous dimensions of all gauge theory operators as functions of 
the 't Hooft coupling $\lambda= g^2 N_c$  or equivalently to find the quantized energy levels 
of a superstring  on $AdS_5 \times S^5$.

Although worldsheet quantum field theories and quantum spin chains seem to be completely different systems,
their solution for large quantum numbers is encoded into the same mathematical structure: 
the asymptotic Bethe Ansatz \cite{8}-\cite{14}. Nevertheless the asymptotic Bethe Ansatz cannot be the final 
answer
to the problem, since it fails above a certain order of perturbation theory on the gauge theory side 
\cite{25}-\cite{36}.
On the gauge theory side, the reason for this breakdown is the appearance of wrapping corrections coming 
from contributions of Feynman graphs which encompass the whole size of the spin chain. On the 
string theory side
these corrections can be explained by contributions coming from virtual particles traveling around the
cylinder.  

Such wrapping corrections in leading order can be taken into account as L\"uscher corrections to the energy of
the superstring sigma model \cite{23,24,25,33}.
Leading order wrapping corrections in $\lambda$ have been calculated for a rich set of operators 
\cite{BJ1,BJ2,Beccaria}.
 The desired all loop exact description of anomalous dimensions (string energies)
is believed to be achieved by
the Thermodynamic Bethe Ansatz (TBA) approach to the superstring sigma model \cite{FrolovTBA0}.
As a first step the TBA equations for the groud state were derived %%@
\cite{stringhyp,KaziY1,KaziY2,FrolovTBA1,OlaszTBA}, then a set of functional equations, 
the so-called Y-system equations 
were proposed to define the exact anomalous dimensions of any physical operator of planar ${\cal N}=4$ 
SYM  \cite{KaziY1}. 
  This Y-system can be
equivalently reformulated as a T-system defined on a "T-shaped fat hook" \cite{KaziY1}.
 The elements of the T-system satisfy discrete Hirota equations \cite{Hirota}. 

In previous works \cite{27K}-\cite{KaziT} discrete Hirota equations were considered as
discrete integrable soliton equations and were solved by means of a chain of B\"acklund 
transformations defining the so-called undressing procedure. In case of $gl(K)$ and $gl(K|M)$
spin chains the T-functions of the Hirota equations correspond to fused transfer
 matrices of the corresponding vertex model \cite{Tfuz,Tsuboi} and in \cite{Wiegmann, KaziT} it was shown that
 the undressing procedure defined by the
 successive application of B\"acklund transformations is equivalent to the nesting procedure
 in the nested Bethe ansatz solution of the spin chains.   

In the present paper we consider discrete Hirota equations as discrete integrable soliton equations
and we solve them by means of a chain of B\"acklund transformations, when they are defined
on a T-shaped fat-hook of arbitrary size.
As a result we get TT-, TQ-, and QQ- relations for
the corresponding T-system comprising the case of AdS/CFT.
  
  The plan of the paper is as follows:
In section 2. we recall the Y-system and the associated T-system for AdS/CFT and the more general
setup which we solve is addressed. In section 3. we present the chain of auto-B\"acklund transformations, 
and the boundary conditions for the hierarchy of Hirota equations.
In section 4. and 5. we derive the TQ-, TT-, and QQ-relations for the T-system defined on a T-shaped
fat-hook. In section 6. the integration of Hirota equations is discussed. In section 7. we apply our
results to the case of AdS/CFT. Finally, section 8 is devoted to summary and discussion. In appendices  
A and B identities and relations skipped from the main text are listed.

\section{Y-, and T-systems for AdS/CFT}

In \cite{KaziY1} a Y-system which yields the exact planar spectrum of 
AdS/CFT was proposed.
The Y-system is a set of functional equations for the functions $Y_{a,s}(u)$ of the spectral
parameter $u$ in such a way that the indices of the Y-functions are defined on a certain domain
of the $(a,s)$ lattice. In case of AdS/CFT the indices take values on the domain represented in figure 1.
The Y-system functional equations take the usual universal form:
\begin{equation} \label{Y}
\frac{Y_{a,s}(u+1)\, Y_{a,s}(u-1)}{Y_{a+1,s}(u) \, Y_{a-1,s}(u)}=
\frac{(1+Y_{a,s+1}(u))\,(1+Y_{a,s-1}(u))}{(1+Y_{a+1,s}(u))\,(1+Y_{a-1,s}(u))}.
\end{equation}
The Y-system of AdS/CFT is equivalent to a T-system which is an integrable discrete dynamics 
on a T-shaped fat-hook (figure 1.) given by the Hirota equations \cite{KaziY1,KaziY2}:
\begin{equation} \label{T}
T_{a,s}(u+1)\,T_{a,s}(u-1)=T_{a+1,s}(u)\,T_{a-1,s}(u)+T_{a,s+1}(u)\,T_{a,s-1}(u),
\end{equation}
where
\begin{equation} \label{YT}
Y_{a,s}(u)=\frac{T_{a,s+1}(u)\, T_{a,s-1}(u)}{T_{a+1,s}(u) \, T_{a-1,s}(u)}, \qquad
1+Y_{a,s}(u)=\frac{T_{a,s}(u+1)\, T_{a,s}(u-1)}{T_{a+1,s}(u) \, T_{a-1,s}(u)}.
\end{equation}

%%%%%%%%%%%%%%%%%%%%%%%%%%%%%%%%%%%%%%%%%%%%%%%%%%%%%%%%%%%%%%%%%%%%%%%%%%
\begin{figure}[htbp]
\begin{center}
\begin{picture}(280,60)  \epsfxsize=70mm
\put(0,15) {\epsfbox{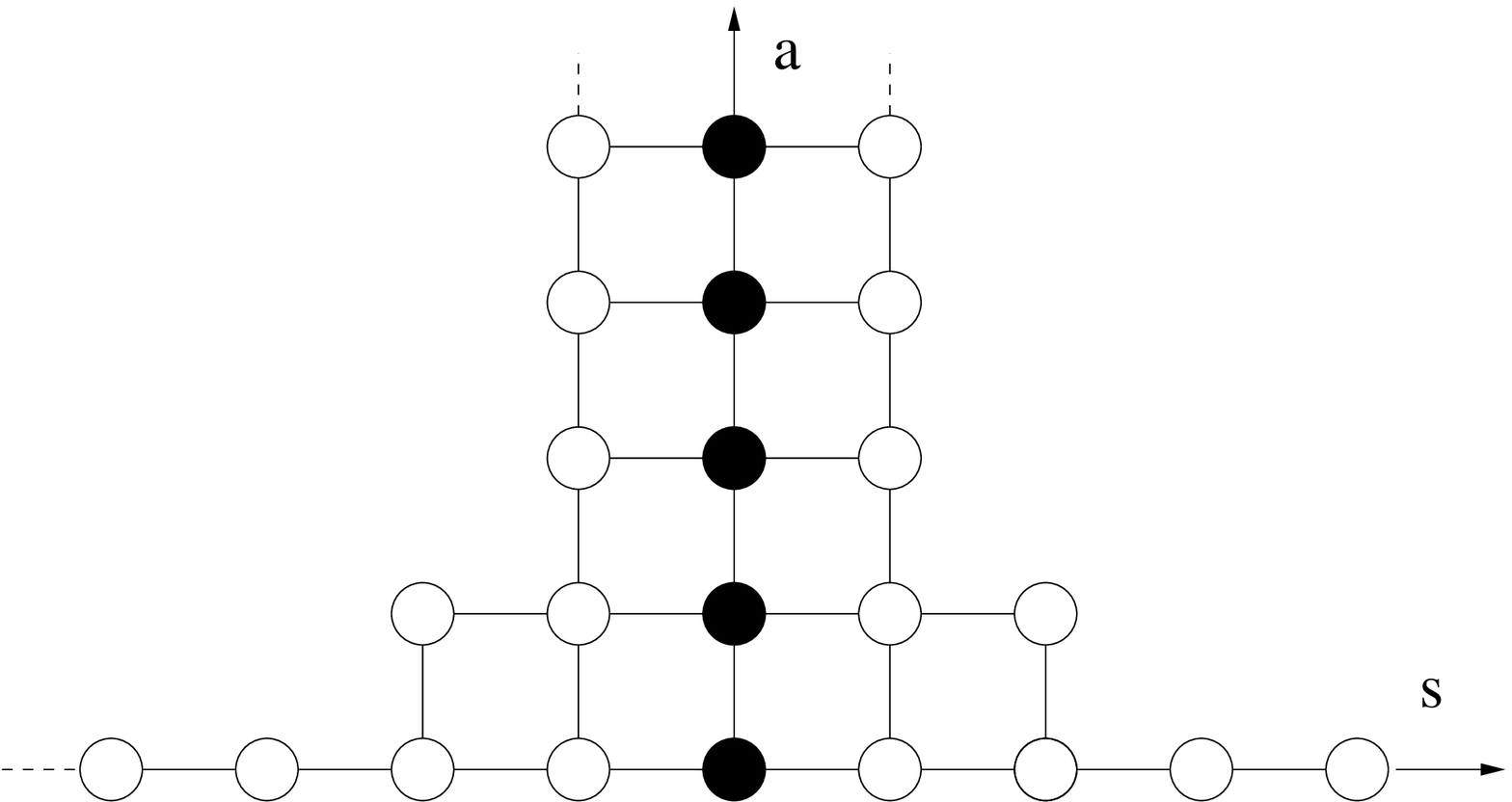}}
\put(100,15) {\epsfbox{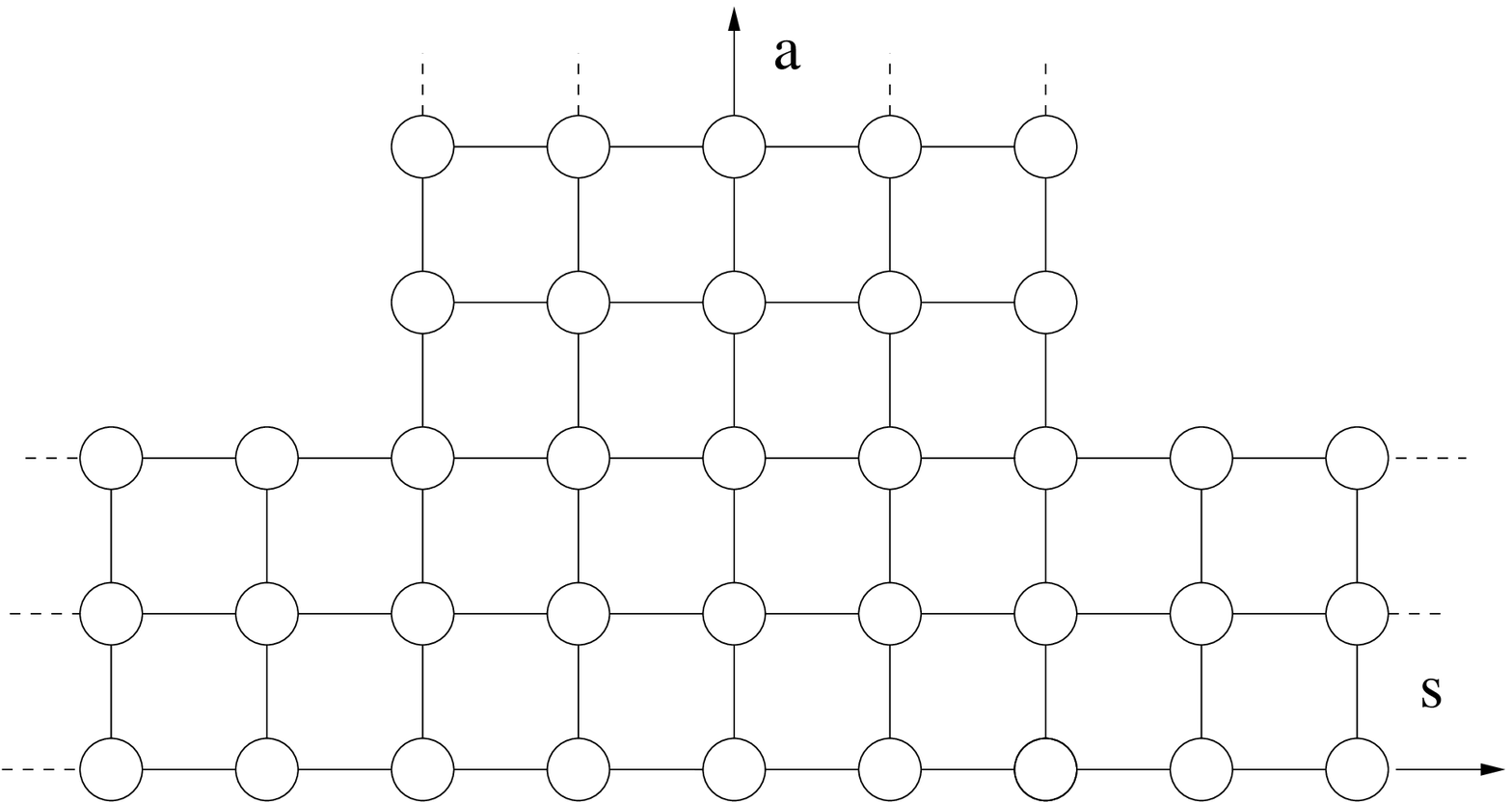}}
\put(0,0) {\parbox{140mm}
{\caption{ \label{6f}\protect {\small
 The domain of definition for the Y-system (on the left) and for the T-system (on the right). }}}}
\end{picture}
\end{center}
\end{figure}
%%%%%%%%%%%%%%%%%%%%%%%%%%%%%%%%%%%%%%%%%%%%%%%%%%%%%%%%%%%%%%%%%%%%%%%%%%%%%%%

Our purpose is to solve the T-system of AdS/CFT, which means that we express the
infinitely many $T_{a,s}(u)$ components of the T-system with a few appropriate 
functions. These functions will be called the Q-functions, and they are nothing but
the functions defining the boundary conditions of the hierarchy of Hirota equations
generated by the B\"acklund transformations defined in the next section. 
%\newline
 
In order to be able to achieve this plan we need to consider T-systems (\ref{T}) on a more
general shaped domain shown in figure 2. The domain can be characterized by
four integer numbers corresponding to its corner point coordinates $(K',-M')$ and $(K,M)$,
where the first coordinate means the "$a$-coordinate", while the second coordinate means the
"$s$-coordinate" of the corner points on the $(a,s)$ lattice. For the sake of simplicity a
T-shaped fat hook with corner point coordinates $(K',-M')$ and $(K,M)$ is denoted by 
$(K',M') \odot (K,M)$ and will be referred as T-shaped fat hook of type $(K',M') \odot (K,M)$.
 The T-functions corresponding to such a domain
will be demoted by $T_{K,M}^{K',M'}(a,s,u)$. The AdS/CFT case corresponds to the special
$K'=M'=K=M=2$ choice for the domain. 
%%%%%%%%%%%%%%%%%%%%%%%%%%%%%%%%%%%%%%%
\begin{figure}[htb]
\begin{flushleft}
%\vskip 10mm
\hskip 15mm
\leavevmode
\epsfxsize=120mm
\epsfbox{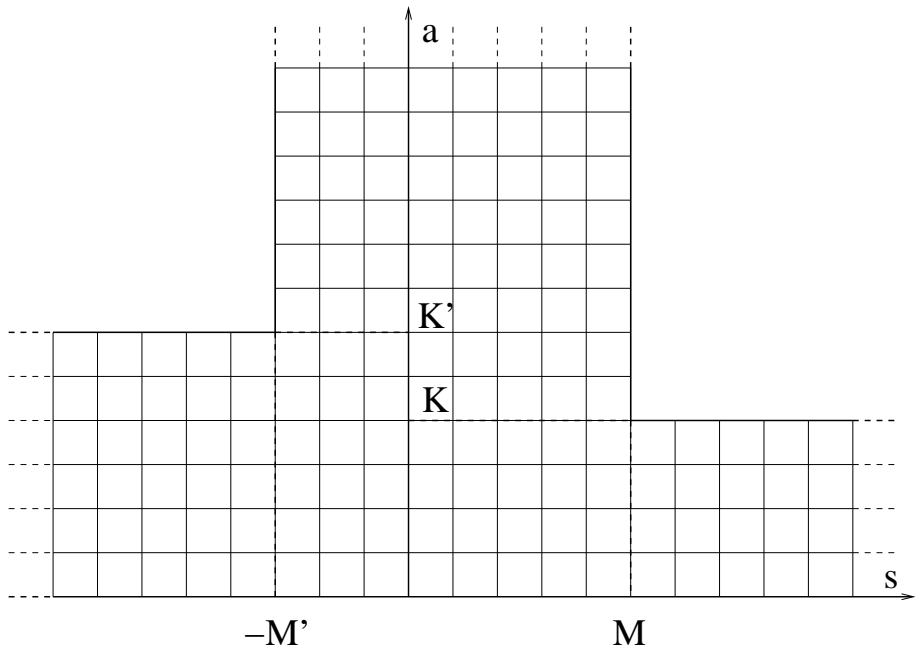}
%\vskip 10mm
\end{flushleft}
\caption{{\footnotesize
General T-shaped fat hook . 
}}
\label{1}
\end{figure}
%%%%%%%%%%%%%%%%%%%%%%%%%%%%%%%%%%%%%%%%%
An important property of the Hirota equations is their invariance with respect to the gauge 
transformations:
\begin{equation} \label{gauge}
T_{a,s}(u) \rightarrow g_1(u+a+s) \, g_2(u+a-s) \, g_3(u-a+s) \, g_4(u-a-s) \, T_{a,s}(u).
\end{equation}
Both Y- and T-systems are sets of functional equations 
which have to be supplemented with certain boundary conditions on the $(a,s)$ lattice. 
Our choice for the boundary conditions of a T-shaped fat hook of type $(K',M') \odot (K,M)$
will be specified in the next section.

\section{Hierarchy of Hirota equations}

This section is devoted to the discussion of the boundary conditions and auto-B\"acklund transformations
of the Hirota equations defined on a T-shaped fat-hook represented in figure 2.
We begin with the  boundary conditions.

\subsection{Boundary conditions for T-shaped fat-hook}

We extend the Hirota equations (\ref{T}) to be valid on the whole $(a,s)$ lattice, in such a way
that the $T_{a,s}(u)$ functions are zero if their indices take values outside of the T-shaped
fat hook of type $(K',M') \odot (K,M)$:
\begin{equation} \label{}
T(a,s,u)=0, \quad \mbox{if:} \quad \mbox{(i)} \,\,\, a<0, \,\,\, \mbox{or} \,\,\, \mbox{(ii)} \,\,\, 
a>K, \, s>M  \,\,\, \mbox{or} \,\,\, \mbox{(iii)} \,\, \, a>K', \, s<-M',   
\end{equation}
where we introduced the notation $T_{a,s}(u)=T(a,s,u)$.
 The boundary values of $T(a,s,u)$ are rather special, because taking the Hirota equations
(\ref{T}) at the boundaries they reduce to discrete d'Alambert equations. For example,
taking (\ref{T}) at $a=0$, the equations reduce to:
\begin{equation} \label{d'Alambert}
T(0,s,u+1)\,T(0,s,u-1)=T(0,s-1,u)\,T(0,s+1,u).
\end{equation}
The solution of these equations can be expressed as a product of two arbitrary functions, 
one is the function of $u-s$ only and the other depends only on $u+s$:
$$T(0,s,u)=f_{-}(u-s)\,f_{+}(u+s). $$
This analysis can be done to all boundary lines of the T-shaped fat-hook and 
finally one gets that the boundary values of $T(a,s,u)$ along the horizontal (vertical)
boundaries are products of two functions, one depending only on $u-s\,\,$, ($u-a$) 
and the other depending only on $u+s \,\,$, ($u+a$). There is an additional constraint that connects
the horizontal and vertical boundary values of $T(a,s,u)$, namely because the corner points $(K,M)$
or $(K',-M')$ are part of both the vertical and horizontal boundary lines, 
the horizontal and vertical boundary values for $T(a,s,u)$ must be equal at these points.
It turns out that the most general boundary conditions satisfying the previous conditions
can be brought by an appropriate gauge transformation (\ref{gauge}) into the following form:
\begin{eqnarray} \label{boundary}  
T(0,s,u) &=& Q^{K',M'}(u+s) \, \tilde{Q}_{K,M}(u-s) \qquad -\infty<s<\infty  \\
T(K',s,u) &=& {\cal Q}_{K,M}^{K'}(u-s+K')\,Q^{0,M'}(u+s-K') \qquad \qquad \qquad s\leq -M' \nonumber \\
T(a,-M',u) &=& (-1)^{M'(a-K')} \, {\cal Q}_{K,M}^{K'}(u+a+M')\,Q^{0,M'}(u-a-M') \quad a\geq K' \nonumber \\
T(K,s,u) &=& {\cal \tilde{Q}}^{K',M'}_{K}(u+s+K)\,\tilde{Q}_{0,M}(u-s-K) \qquad \qquad \qquad s\geq M \nonumber %%@
\\
T(a,M,u) &=& (-1)^{M(a-K)} \, {\cal \tilde{Q}}^{K',M'}_{K}(u+a+M)\,\tilde{Q}_{0,M}(u-a-M) \quad a\geq K. %%@
\nonumber
\end{eqnarray}
The functions appearing on the right hand side of (\ref{boundary}) characterize the boundary conditions
for $T(a,s,u)$.
In the present paper our purpose is to solve the Hirota equations (\ref{T}) when boundary conditions
(\ref{boundary}) are imposed.

Our strategy of solving Hirota equations agrees with the method worked out in \cite{Wiegmann}.
Namely, we consider the Hirota equations as discrete integrable soliton equations, and by
giving their appropriate auto-B\"acklund transformations  we can construct 
a hierarchy of Hirota equations, in such a way that two neighboring levels of the hierarchy
are connected by a B\"acklund transformation. The hierarchy of Hirota equations consists of
Hirota equations defined on T-shaped fat hooks with different corner point coordinates.
Two members of the hierarchy are called neighboring, when only one of 
the four corner point coordinates of their T-shaped fat hooks differ by $1$.
(i.e. one of the parameters $K',M',K,M$ differ by $1$).
 The existence of such a hierarchy follows from the classical integrability of the Hirota equations.
With the help of the B\"acklund transformations one can decrease $K,M$ or $K',M'$ by 1, "undressing" 
step by step the original $(K',M')\odot
(K,M)$ problem to the trivial $(0,0)\odot(0,0)$ or $(0,-m)\odot(0,m)$ problem.
Then taking the "inverse" of the undressing procedure the solution for the Hirota equations can be
built up from the trivial $(0,0)\odot(0,0)$ problem. This procedure will be explained  in section 6.

The "undressing" procedure was shown to be equivalent to algebraic Bethe
ansatz in case of $gl(K)$ \cite{Zabrodin,Wiegmann} and $gl(K|M)$ \cite{KaziT} spin chains. In the $gl(K)$ case 
the domain of non-zero $T(a,s,u)$ is a semi-infinite strip of height K \cite{Wiegmann}, while 
in case of $gl(K|M)$ super algebras this domain is a fat-hook with corner point
coordinates $(K,M)$ \cite{KaziT}. 

In the next subsection we define the auto-B\"acklund transformations of the Hirota equations
which provide us with a hierarchy of Hirota equations. 

\subsection{B\"acklund transformations}

In this subsection we define 4 auto-B\"acklund transformations of the Hirota equations each
corresponding to shifting one of the coordinates $K,M,K',M'$ by 1 unit. We use the
terminology of left and right B\"acklund transformations in the following sense:
if a B\"acklund transformation decreases the value of $K$ or $M$ by 1 we call it right
B\"acklund transformation, and if it decreases the value of $K'$ or $M'$ we call it left
B\"acklund transformation. Right B\"acklund transformations change the corner point coordinates
of the T-shaped fat-hook on the right hand side and leave the left hand side intact, while
the left B\"acklund transformations change the T-shaped fat-hook on the left hand side and leave
the right hand side intact.
 The right B\"acklund transformations take exactly the same form as those  
defined for the $gl(K|M)$ super spin chains \cite{KaziT}. Let us call the B\"acklund transformed functions
$F(a,s,u)$, then the defining equations of the right B\"acklund transformations read as \cite{KaziT}:
{\normalsize \begin{eqnarray}
T(a+1,s,u)\,F(a,s,u+1)-T(a,s,u+1)\, F(a+1,s,u) &=& \nonumber \\
T(a+1,s-1,u+1) \, F(a,s+1,u), \nonumber \\
T(a,s+1,u+1)\,F(a,s,u)-T(a,s,u)\, F(a,s+1,u+1) &=& \nonumber \\
T(a+1,s,u+1) \, F(a-1,s+1,u).  \label{TF}
\end{eqnarray} }
or equivalently after some simple manipulations described in \cite{KaziT} it can be written as a single matrix 
equation
\begin{equation} \label{MTF}
{\mathbb T} \, {\bf F}=0, 
\end{equation}
where the matrix ${\mathbb T}$ is given by:
 \begin{equation}
{\mathbb T} = \left(\begin{array}{cccc}
0 & T(a,s,u-1) & -T(a,s+1,u) & T(a+1,s,u) \\
-T(a,s,u-1) & 0 & T(a-1,s,u) & T(a,s-1,u) \\
T(a,s+1,u) & -T(a-1,s,u) & 0 & -T(a,s,u+1) \\
-T(a+1,s,u) & -T(a,s-1,u) & T(a,s,u+1) & 0
\end{array}\right)
\end{equation}
and the vector ${\bf F}$ reads as:
\begin{equation}
{\bf F}= \left(\begin{array}{c}
F(a-1,s,u) \\
F(a,s+1,u) \\
F(a,s,u-1) \\
F(a-1,s+1,u-1) 
\end{array} \right).
\end{equation} 
It can be seen that the right B\"acklund transformations are homogeneous linear equations
for $F(a,s,u)$ which have nontrivial solutions only if the determinant of the matrix ${\mathbb T}$
vanishes. In fact this happens because $T(a,s,u)$ satisfies the Hirota equations, and it turns out
that the rank of the matrix ${\mathbb T}$ is 2, so there are 2 linearly independent solutions
of the linear problem (\ref{MTF}).

There is a duality between $T(a,s,u)$ and $F(a,s, u)$ \cite{KaziT}, namely one can interchange the
roles of $T$ and $F$ and consider (\ref{TF}) as an over-determined system of
linear equations for $T$ with coefficients $F$. Their compatibility equation
says that $F(a,s,u)$ satisfies the Hirota equations (\ref{T}) as well.
However it can be shown that $F$ can not be defined on the same T-shaped fat-hook as $T$.
Careful analysis of the boundary conditions for $F$ compatible with (\ref{TF}) says
that $F$ must be defined either on a T-shaped fat-hook of type $(K',M')\odot (K-1,M)$
or on a T-shaped fat-hook of type $(K',M')\odot (K,M+1)$. These two possible choices for
the domain of definition correspond to the 2 linearly independent solutions of (\ref{MTF}).
We want such right B\"acklund transformations which decrease the value of $K$ or $M$ by 1 unit.
The first ($K$ decreasing) right B\"acklund transformation (BT1) is defined by equations (\ref{TF}),
 while the second ($M$ decreasing) right 
transformation can be obtained from (\ref{TF}) by interchanging the role of $T$ and $F$
\cite{KaziT}. We denote it (BT2) for short, and it reads:
\begin{eqnarray}
F^*(a+1,s,u)\,T(a,s,u+1)-F^*(a,s,u+1)\, T(a+1,s,u) &=& \nonumber \\
F^*(a+1,s-1,u+1) \, T(a,s+1,u), \nonumber \\
F^*(a,s+1,u+1)\,T(a,s,u)-F^*(a,s,u)\, T(a,s+1,u+1) &=& \nonumber \\
F^*(a+1,s,u+1) \, T(a-1,s+1,u).  \label{TF*}
\end{eqnarray}
 The function $F^*(a,s,u)$ is defined on a T-shaped fat-hook of type $(K',M')\odot (K,M-1)$.
With the help of the (BT1) and (BT2) B\"acklund transformations one can decrease the size
of the T-shaped fat-hook on its right hand side by 1 unit in either vertical or 
horizontal direction. 

 In order that the T-shaped fat-hook can be reduced to a trivial domain one needs another set of 
B\"acklund transformations which decrease the size of the T-shaped fat-hook on its left hand side. 
These left B\"acklund transformations can be easily created from (\ref{TF}) after performing
a reflection transformation in the $s$ variable. Denoting the left B\"acklund transformed function 
$\tilde{F}(a,s,u)$ they read as:
{\normalsize \begin{eqnarray}
T(a+1,s,u)\,\tilde{F}(a,s,u+1)-T(a,s,u+1)\, \tilde{F}(a+1,s,u) &=& \nonumber \\
T(a+1,s+1,u+1) \, \tilde{F}(a,s-1,u), \nonumber \\
T(a,s-1,u+1)\,\tilde{F}(a,s,u)-T(a,s,u)\, \tilde{F}(a,s-1,u+1) &=& \nonumber \\
T(a+1,s,u+1) \, \tilde{F}(a-1,s-1,u).  \label{TFL}
\end{eqnarray} }
or equivalently it can be written as a single matrix equation: 
\begin{equation} \label{MTFL}
\tilde{{\mathbb T}} \, {\bf \tilde{F}}=0, 
\end{equation}
where the matrix $\tilde{{\mathbb T}}$ is given by:
 \begin{equation}
\tilde{{\mathbb T}} = \left(\begin{array}{cccc}
0 & T(a,s,u-1) & -T(a,s-1,u) & T(a+1,s,u) \\
-T(a,s,u-1) & 0 & T(a-1,s,u) & T(a,s+1,u) \\
T(a,s-1,u) & -T(a-1,s,u) & 0 & -T(a,s,u+1) \\
-T(a+1,s,u) & -T(a,s+1,u) & T(a,s,u+1) & 0
\end{array}\right)
\end{equation}
and the vector ${\bf \tilde{F}}$ reads as:
\begin{equation}
{\bf \tilde{F}}= \left(\begin{array}{c}
\tilde{F}(a-1,s,u) \\
\tilde{F}(a,s-1,u) \\
\tilde{F}(a,s,u-1) \\
\tilde{F}(a-1,s-1,u-1) 
\end{array} \right).
\end{equation} 
Following similar train of thoughts as in case of the right B\"acklund transformations it can be shown
that (\ref{MTFL}) has also two linearly independent solutions which satisfy the Hirota equations.
They correspond to T-shaped fat hooks of type $(K'-1,M') \odot (K,M)$ or  $(K',M'+1) \odot (K,M)$.
Then the first ($K'$ decreasing) left B\"acklund transformation ($\overline{BT1}$) is defined 
by equations (\ref{TFL}), while the second ($M'$ decreasing)
 transformation can be obtained from (\ref{TFL}) by interchanging the role of $T$ and $\tilde{F}$.
 We call this second left B\"acklund transformation ($\overline{BT2}$), and it reads:
\begin{eqnarray}
\tilde{F}^*(a+1,s,u)\,T(a,s,u+1)-\tilde{F}^*(a,s,u+1)\, T(a+1,s,u) &=& \nonumber \\
\tilde{F}^*(a+1,s+1,u+1) \, T(a,s-1,u), \nonumber \\
\tilde{F}^*(a,s-1,u+1)\,T(a,s,u)-\tilde{F}^*(a,s,u)\, T(a,s-1,u+1) &=& \nonumber \\
\tilde{F}^*(a+1,s,u+1) \, T(a-1,s-1,u).  \label{TF*L}
\end{eqnarray}
 The function $F^*(a,s,u)$ is defined on a T-shaped fat-hook of type $(K',M'-1)\odot (K,M)$.
With the help of the ($\overline{BT1}$) and ($\overline{BT2}$) B\"acklund transformations one can decrease the %%@
size
of the T-shaped fat-hook on its left hand side by 1 unit in either vertical or 
horizontal direction. The actions of (BT1), (BT2), ($\overline{BT1}$) and ($\overline{BT2}$) B\"acklund %%@
transformations
are depicted in figure 3.
%%%%%%%%%%%%%%%%%%%%%%%%%%%%%%%%%%%%%%%
\begin{figure}[htb]
\begin{flushleft}
%\vskip 10mm
\hskip 15mm
\leavevmode
\epsfxsize=120mm
\epsfbox{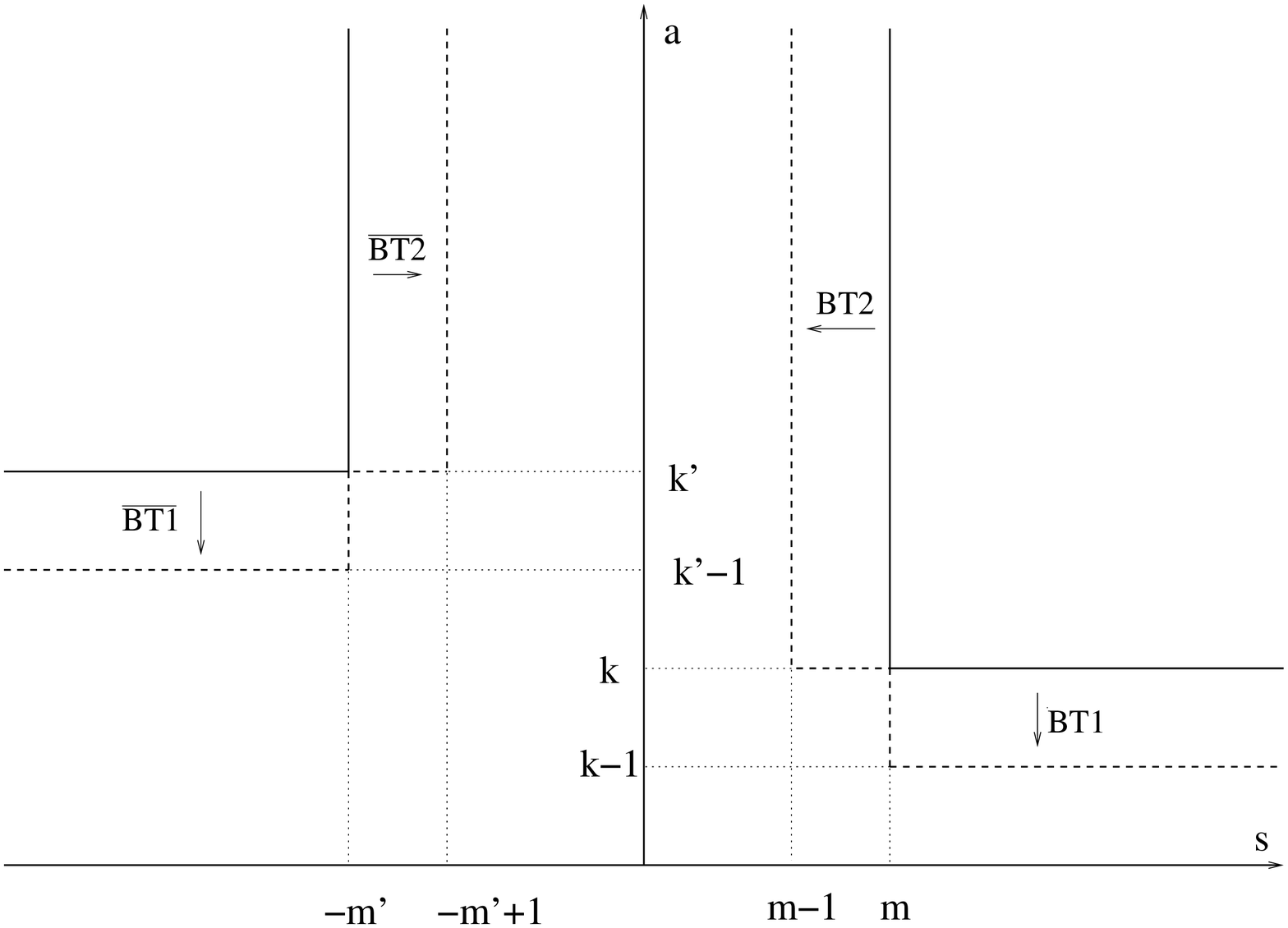}
%\vskip 10mm
\end{flushleft}
\caption{{\footnotesize
Schematic representation of the actions of (BT1), (BT2), ($\overline{BT1}$) and ($\overline{BT1}$) B\"acklund %%@
transformations. 
}}
\label{3}
\end{figure}
%%%%%%%%%%%%%%%%%%%%%%%%%%%%%%%%%%%%%%%%%

\subsection{Boundary conditions for the hierarchy of Hirota equations}

First of all we introduce the notation $T_{k,m}^{k',m'}(a,s,u)$ for T-functions, which
solve the Hirota equations defined on a T-shaped fat-hook of type $(k',m')\odot(k,m)$.
%%%%%%%%%%%%%%%%%%%%%%%%%%%%%%%%%%%%%%%
\begin{figure}[htb]
\begin{flushleft}
%\vskip 10mm
\hskip 15mm
\leavevmode
\epsfxsize=120mm
\epsfbox{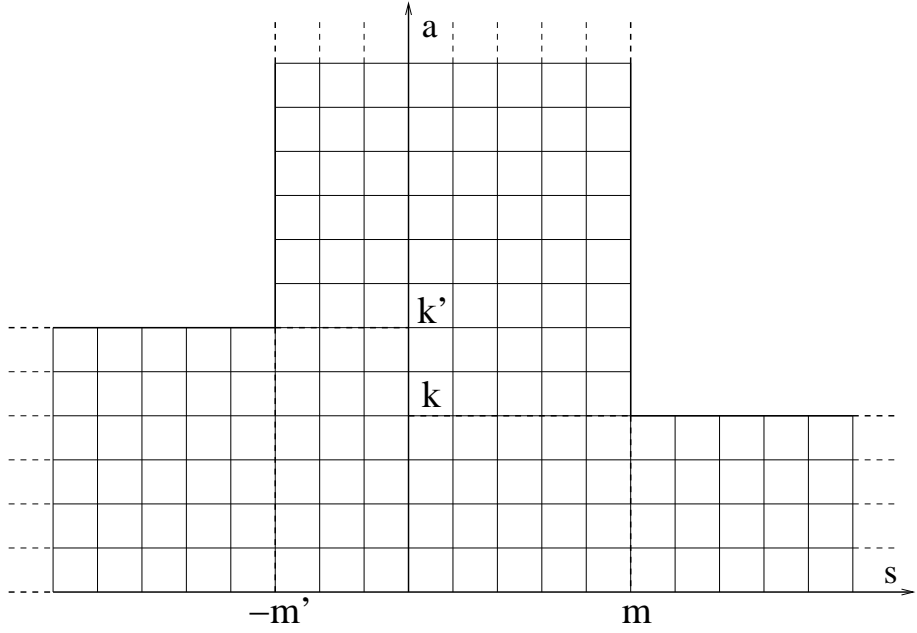}
%\vskip 10mm
\end{flushleft}
\caption{{\footnotesize
Domain of definition for T-functions $T_{k,m}^{k',m'}(a,s,u)$. 
}}
\label{2}
\end{figure}
%%%%%%%%%%%%%%%%%%%%%%%%%%%%%%%%%%%%%%%%%
With the help of this new notation the matrix form of right B\"acklund transformations
 (\ref{TF})-(\ref{TF*}) can be written as:
\begin{equation} \label{MTFkm}
\mbox{BT1} \quad \mbox{case:} \quad
{\mathbb T}_{k,m}^{k',m'} \, {\bf T}_{k-1,m}^{k',m'}=0, \qquad \qquad
\mbox{BT2} \quad \mbox{case:} \quad
{\mathbb T}_{k,m}^{k',m'} \, {\bf T}_{k,m+1}^{k',m'}=0, 
\end{equation}
where the matrix ${\mathbb T}_{k,m}^{k',m'}$ is given by:
 { \footnotesize \begin{equation}
{\mathbb T}_{k,m}^{k',m'} = \left(\begin{array}{cccc}
0 & T_{k,m}^{k',m'}(a,s,u-1) & -T_{k,m}^{k',m'}(a,s+1,u) & T_{k,m}^{k',m'}(a+1,s,u) \\
-T_{k,m}^{k',m'}(a,s,u-1) & 0 & T_{k,m}^{k',m'}(a-1,s,u) & T_{k,m}^{k',m'}(a,s-1,u) \\
T_{k,m}^{k',m'}(a,s+1,u) & -T_{k,m}^{k',m'}(a-1,s,u) & 0 & -T_{k,m}^{k',m'}(a,s,u+1) \\
-T_{k,m}^{k',m'}(a+1,s,u) & -T_{k,m}^{k',m'}(a,s-1,u) & T_{k,m}^{k',m'}(a,s,u+1) & 0
\end{array}\right)
\end{equation} }
and the vector ${\bf T}_{k,m}^{k',m'}$ reads as:
\begin{equation}
{\bf T}_{k,m}^{k',m'}= \left(\begin{array}{c}
T_{k,m}^{k',m'}(a-1,s,u) \\
T_{k,m}^{k',m'}(a,s+1,u) \\
T_{k,m}^{k',m'}(a,s,u-1) \\
T_{k,m}^{k',m'}(a-1,s+1,u-1) 
\end{array} \right).
\end{equation} 
The matrix form of left B\"acklund transformations
 (\ref{TFL})-(\ref{TF*L}) can be written as:
\begin{equation} \label{MTFk'm'}
\overline{BT1} \quad \mbox{case:} \quad
\tilde{{\mathbb T}}_{k,m}^{k',m'} \, {\bf \tilde{T}}_{k,m}^{k'-1,m'}=0, \qquad  \qquad
\overline{BT2} \quad \mbox{case:} \quad
\tilde{{\mathbb T}}_{k,m}^{k',m'} \, {\bf \tilde{T}}_{k,m}^{k',m'+1}=0, 
\end{equation}
where the matrix $\tilde{{\mathbb T}}_{k,m}^{k',m'}$ is given by:
 { \footnotesize \begin{equation}
\tilde{{\mathbb T}}_{k,m}^{k',m'} = \left(\begin{array}{cccc}
0 & T_{k,m}^{k',m'}(a,s,u-1) & -T_{k,m}^{k',m'}(a,s-1,u) & T_{k,m}^{k',m'}(a+1,s,u) \\
-T_{k,m}^{k',m'}(a,s,u-1) & 0 & T_{k,m}^{k',m'}(a-1,s,u) & T_{k,m}^{k',m'}(a,s+1,u) \\
T_{k,m}^{k',m'}(a,s-1,u) & -T_{k,m}^{k',m'}(a-1,s,u) & 0 & -T_{k,m}^{k',m'}(a,s,u+1) \\
-T_{k,m}^{k',m'}(a+1,s,u) & -T_{k,m}^{k',m'}(a,s+1,u) & T_{k,m}^{k',m'}(a,s,u+1) & 0
\end{array}\right),
\end{equation} }
and the vector ${\bf \tilde{T}}_{k,m}^{k',m'}$ reads as:
\begin{equation}
{\bf \tilde{T}}_{k,m}^{k',m'}= \left(\begin{array}{c}
T_{k,m}^{k',m'}(a-1,s,u) \\
T_{k,m}^{k',m'}(a,s-1,u) \\
T_{k,m}^{k',m'}(a,s,u-1) \\
T_{k,m}^{k',m'}(a-1,s-1,u-1) 
\end{array} \right).
\end{equation} 
The (\ref{MTFkm}) and (\ref{MTFk'm'}) B\"acklund transformations relate the boundary conditions of
$T_{k,m}^{k',m'}(a,s,u)$ at different values of their indices. If one starts the undressing
procedure from $T_{K,M}^{K',M'}(a,s,u)$ with boundary conditions given by (\ref{boundary}),
then all the $T_{k,m}^{k',m'}(a,s,u)$ functions of the hierarchy of Hirota equations have structurally
 the same type of boundary conditions: 
\begin{eqnarray} \label{boundarykm}  
T_{k,m}^{k',m'}(0,s,u) &=& Q^{k',m'}(u+s) \, \tilde{Q}_{k,m}(u-s) \qquad -\infty<s<\infty  \\
T_{k,m}^{k',m'}(k',s,u) &=& {\cal Q}_{k,m}^{k'}(u-s+k')\,Q^{0,m'}(u+s-k') \qquad \qquad \qquad s\leq -m' %%@
\nonumber \\
T_{k,m}^{k',m'}(a,-m',u) &=& (-1)^{m'(a-k')} \, {\cal Q}_{k,m}^{k'}(u+a+m')\,Q^{0,m'}(u-a-m') \quad a\geq k' %%@
\nonumber \\
T_{k,m}^{k',m'}(k,s,u) &=& {\cal \tilde{Q}}^{k',m'}_{k}(u+s+k)\,\tilde{Q}_{0,m}(u-s-k) \qquad \qquad \qquad %%@
s\geq m \nonumber \\
T_{k,m}^{k',m'}(a,m,u) &=& (-1)^{m(a-k)} \, {\cal \tilde{Q}}^{k',m'}_{k}(u+a+m)\,\tilde{Q}_{0,m}(u-a-m) \quad %%@
a\geq k. \nonumber
\end{eqnarray}
The dependence of the boundary functions on their indices is dictated by the fact that one gets all the
$T_{k,m}^{k',m'}(a,s,u)$ functions from $T_{K,M}^{K',M'}(a,s,u)$ by successive application of 
 B\"acklund transformations (\ref{MTFkm}) and (\ref{MTFk'm'}). 
 
At the present stage of the presentation the boundary functions on the right hand side of (\ref{boundarykm})
seem to be arbitrary functions of $u$. But this is not the case. There are certain relations among them.

The first class of relations comes from such degenerate cases, when certain parts of the boundary lines
of the T-shaped fat hook coincide. For example, for the special choice of $m=-m'$, there is a common part 
of the vertical left and right boundary lines of the T-shaped fat-hook (see figure 5.). This degeneration %%@
connects
the boundary functions on the left and right hand sides of the T-shaped fat hook. Relations coming
from such coincidences of boundary lines are listed in appendix A.

The second class of relations among boundary functions of (\ref{boundarykm}) are given by the so-called
QQ-relations, which will be the subject of section 5.

%%%%%%%%%%%%%%%%%%%%%%%%%%%%%%%%%%%%%%%
\begin{figure}[htb]
\begin{flushleft}
%\vskip 10mm
\hskip 15mm
\leavevmode
\epsfxsize=100mm
\epsfbox{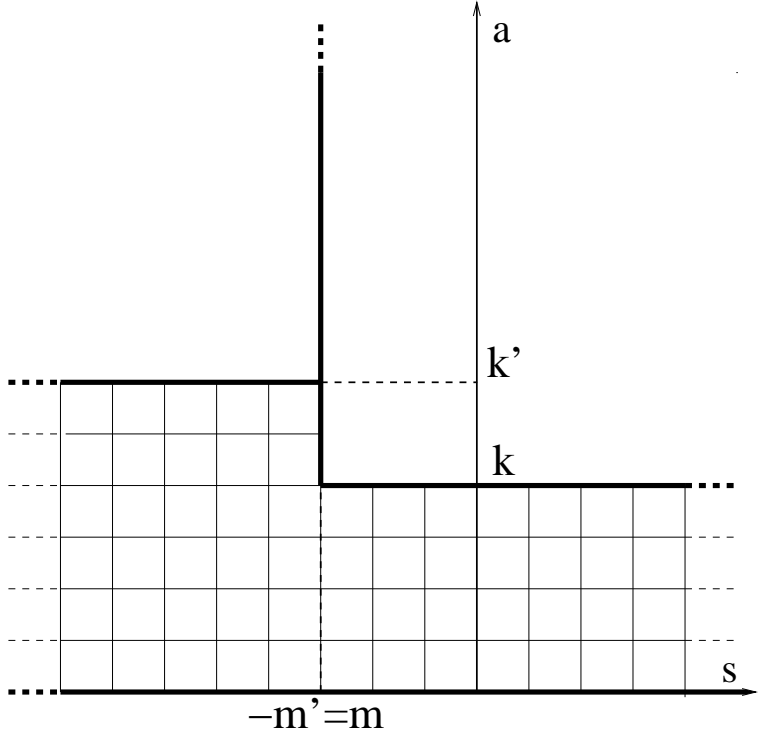}
%\vskip 10mm
\end{flushleft}
\caption{{\footnotesize
Degenerate domain corresponding to the choice $m=-m'$. 
}}
\label{5a}
\end{figure}
%%%%%%%%%%%%%%%%%%%%%%%%%%%%%%%%%%%%%%%%%
%%%%%%%%%%%%%%%%%%%%%%

\section{TQ-relations}

This short section summarizes the TQ-relations for the Hirota equations defined on the T-shaped
fat-hook of type $(K',M') \odot (K,M)$. In our case generalized Baxter equations or TQ-relations
are  bilinear equations between the functions $T_{k,m}^{k',m'}(a,s,u)$ and $Q^{k',m'}(u)$ or
$\tilde{Q}_{k,m}(u)$, because these bilinear relations can be considered as generalizations of
the TQ-relations of the simplest and well known case of the $SU(2)$ spin chain \cite{TQ}.  
As it was shown in \cite{Zabrodin,Wiegmann} the Q-functions of the spin chains can be identified with the
 functions characterizing the boundary conditions of the domain of definition of the T-functions.
The set of functions one can identify as the $Q$-functions of the problem are the  
$Q^{k',m'}(u)$ and $\tilde{Q}_{k,m}(u)$ functions. In this case the TQ-relations  
 can be obtained by taking the (\ref{MTFkm}) (BT1), (BT2),
 and (\ref{MTFk'm'}), ($\overline{BT1}$),
 ($\overline{BT2}$)   B\"acklund transformations along the boundary line $(0,s)$. 
 TQ-relations coming from (BT1) and (BT2) B\"acklund transformations read as follows:
\begin{eqnarray} \label{TQBT1}
T_{k,m}^{k',m'}(1,s,u) \,\tilde{Q}_{k-1,m}(u-s+1)-T_{k-1,m}^{k',m'}(1,s,u) \,\tilde{Q}_{k,m}(u-s+1)&=&
\nonumber \\
T_{k,m}^{k',m'}(1,s-1,u+1) \,\tilde{Q}_{k-1,m}(u-s-1) \nonumber \\
T_{k,m-1}^{k',m'}(1,s,u) \,\tilde{Q}_{k,m}(u-s+1)-T_{k,m}^{k',m'}(1,s,u) \,\tilde{Q}_{k,m-1}(u-s+1)&=&
\nonumber \\
T_{k,m-1}^{k',m'}(1,s-1,u+1) \,\tilde{Q}_{k,m}(u-s-1),
\end{eqnarray}
while TQ-relations coming from ($\overline{BT1}$), ($\overline{BT2}$) B\"acklund transformations
are as follows: 
\begin{eqnarray} \label{TQBT1L}
T_{k,m}^{k',m'}(1,s,u) \,Q^{k'-1,m'}(u+s+1)-T_{k,m}^{k'-1,m'}(1,s,u) \,Q^{k',m'}(u+s+1)&=&
\nonumber \\
T_{k,m}^{k',m'}(1,s+1,u+1) \,Q^{k'-1,m'}(u+s-1) \nonumber \\
T_{k,m}^{k',m'-1}(1,s,u) \,Q^{k',m'}(u+s+1)-T_{k,m}^{k',m'}(1,s,u) \,Q^{k',m'-1}(u+s+1)&=&
\nonumber \\
T_{k,m}^{k',m'-1}(1,s+1,u+1) \,Q^{k',m'}(u+s-1).
\end{eqnarray}
In references \cite{Wiegmann} and \cite{KaziT} these TQ-relations were used to construct an operator
 generating series for the T-functions. The generating series was factorized into an ordered product 
of first order difference operators with coefficients being ratios of the $Q$-functions. 
Here we skip the construction of such generating series
and we will discuss the problem of expressing the T-functions in terms of the $Q^{k',m'}(u)$ and
$\tilde{Q}_{k,m}(u)$  functions later in section 6. 
  
  Closing this section we have to mention that the (\ref{MTFkm}) and (\ref{MTFk'm'}) B\"acklund
transformations written along the right and left vertical boundary lines imply additional TQ type
bilinear relations between the T-functions and the boundary characterizing functions. They are listed
in Appendix B. 

\section{TT- and QQ-relations}

In this section we derive bilinear equations for the functions $T_{k,m}^{k',m'}(a,s,u)$ in which
either both indices $k,m$ or both indices $k',m'$ undergo shifts by $\pm 1$. These equations are
also of Hirota type and we call them TT-relations. The special cases  (\ref{QQtilde}) and (\ref{QQv}) 
are particularly important.
They provide bilinear relations for the $Q$ and $\tilde{Q}$ functions which we call
QQ-relations for traditional reasons.   

All these bilinear TT-relations follow from the (BT1), (BT2) and ($\overline{BT1}$), 
($\overline{BT2}$) B\"acklund transformations with shifts in $k$ and $m$ or in $k'$ and $m'$.Following the %%@
lines of \cite{KaziT} it can be shown that the first rows of (\ref{MTFkm}) can be rewritten as  
\begin{equation} \label{Apszi}
\Psi_{k-1,m}^{k',m'}(a,s,u)=\hat{{\cal A}}_{k,m}^{k',m'}(a,s,u) \, \Psi_{k,m}^{k',m'}(a,s,u),
\end{equation}
\begin{equation} \label{Bpszi}
\Psi_{k,m+1}^{k',m'}(a,s,u)=\hat{{\cal B}}_{k,m}^{k',m'}(a,s,u)\, \Psi_{k,m}^{k',m'}(a,s,u),
\end{equation}
where
\begin{equation} \label{pszi}
\Psi_{k,m}^{k',m'}(a,s,u)=\frac{T_{k,m}^{k',m'}(a+1,s,u)}{T_{k,m}^{k',m'}(a,s+1,u)},
\end{equation}
and $\hat{{\cal A}}_{k,m}^{k',m'}(a,s,u)$  and $\hat{{\cal B}}_{k,m}^{k',m'}(a,s,u)$
are operators acting on the functions $\Psi_{k,m}^{k',m'}(a,s,u)$
\begin{equation} \label{A}
\hat{{\cal A}}_{k,m}^{k',m'}(a,s,u)=
\frac{T_{k-1,m}^{k',m'}(a,s,u+1) \,T_{k,m}^{k',m'}(a,s+1,u)}
{T_{k-1,m}^{k',m'}(a,s+1,u) \,T_{k,m}^{k',m'}(a,s,u+1)}-e^{\partial_u-\partial_s},                  
\end{equation}
\begin{equation} \label{Be}
\hat{{\cal B}}_{k,m}^{k',m'}(a,s,u)=
\frac{T_{k,m+1}^{k',m'}(a,s,u+1) \,T_{k,m}^{k',m'}(a,s+1,u)}
{T_{k,m+1}^{k',m'}(a,s+1,u) \,T_{k,m}^{k',m'}(a,s,u+1)}-e^{\partial_u-\partial_s}.                  
\end{equation}
 The second rows of (\ref{MTFkm}) can be similarly rewritten:
\begin{equation} \label{CFi}
\Phi_{k+1,m}^{k',m'}(a,s,u)=\hat{{\cal C}}_{k,m}^{k',m'}(a,s,u) \, \Phi_{k,m}^{k',m'}(a,s,u),
\end{equation}
\begin{equation} \label{DFi}
\Phi_{k,m-1}^{k',m'}(a,s,u)=\hat{{\cal D}}_{k,m}^{k',m'}(a,s,u) \, \Phi_{k,m}^{k',m'}(a,s,u),
\end{equation}
where 
\begin{equation} \label{Fi}
\Phi_{k,m}^{k',m'}(a,s,u)=\frac{T_{k,m}^{k',m'}(a,s+1,u)}{T_{k,m}^{k',m'}(a+1,s,u)} 
\end{equation}
and $\hat{{\cal C}}_{k,m}^{k',m'}(a,s,u)$  and $\hat{{\cal D}}_{k,m}^{k',m'}(a,s,u)$
are operators acting on the functions $\Phi_{k,m}^{k',m'}(a,s,u)$:
\begin{equation} \label{C}
\hat{{\cal C}}_{k,m}^{k',m'}(a,s,u)=
 \frac{T_{k+1,m}^{k',m'}(a,s,u-1) \,T_{k,m}^{k',m'}(a+1,s,u)}
{T_{k+1,m}^{k',m'}(a+1,s,u) \,T_{k,m}^{k',m'}(a,s,u-1)}+e^{-\partial_u-\partial_a},
\end{equation}
\begin{equation} \label{D}
\hat{{\cal D}}_{k,m}^{k',m'}(a,s,u)=
 \frac{T_{k,m-1}^{k',m'}(a,s,u-1) \,T_{k,m}^{k',m'}(a+1,s,u)}
{T_{k,m-1}^{k',m'}(a+1,s,u) \,T_{k,m}^{k',m'}(a,s,u-1)}+e^{-\partial_u-\partial_a}.
\end{equation}
In this form the (BT1) and (BT2) B\"acklund transformations appear as linear problems for the difference
operators (\ref{A}), (\ref{Be}), (\ref{C}) and (\ref{D}) with particular solutions (\ref{pszi})
and (\ref{Fi}).

Equations (\ref{Apszi}) and (\ref{Bpszi}) hold for any $k,m,k',m'$, so the function
$\Psi_{k-1,m+1}^{k',m'}(a,s,u)$ can be represented in two different ways 
by subsequent action of operators (\ref{A}), (\ref{Be}):
\begin{eqnarray}
\Psi_{k-1,m+1}^{k',m'}(a,s,u)\equiv\hat{{\cal B}}_{k-1,m}^{k',m'}(a,s,u)\,
\hat{{\cal A}}_{k,m}^{k',m'}(a,s,u) \, \Psi_{k,m}^{k',m'}(a,s,u)= \nonumber \\
\hat{{\cal A}}_{k,m+1}^{k',m'}(a,s,u)\,
\hat{{\cal B}}_{k,m}^{k',m'}(a,s,u) \, \Psi_{k,m}^{k',m'}(a,s,u). \label{AB}
\end{eqnarray}
Expanding both sides of (\ref{AB}) and canceling the identical terms, one ends up with a non-trivial
relation connecting the T-functions with different $k$ and $m$. It takes the form
\begin{eqnarray} 
\frac{T_{k,m}^{k',m'}(a,s+1,u) \, T_{k+1,m+1}^{k',m'}(a,s,u+1)-T_{k,m}^{k',m'}(a,s,u+1) \,
T_{k+1,m+1}^{k',m'}(a,s+1,u)}{T_{k+1,m}^{k',m'}(a,s,u+1) \, T_{k,m+1}^{k',m'}(a,s+1,u)}= \nonumber \\
f_{k,m}^{k',m'}(a,u+s), \label{TTf}
\end{eqnarray}
where $f_{k,m}^{k',m'}(a,u+s)$ is an arbitrary function of its variables. Comparing (\ref{TTf}) with 
a similar equation
obtained in the same way from the other pair of linear problems (\ref{CFi}) and (\ref{DFi}), one can see that
$f_{k,m}^{k',m'}$ depends only on the combination $u+s-a$ as well as on its indices. This function can be
 fixed by taking $s=m$ in (\ref{TTf}) and exploiting the boundary conditions (\ref{boundarykm}). 
This fixes the function $f_{k,m}^{k',m'}(a,u+s)$ to be $1$.

Following a similar train of thoughts such TT-relations can be derived from the pair of linear problems 
(\ref{CFi}) and (\ref{DFi}). Thus the two main TT-relations coming from the right B\"acklund transformations
(\ref{MTFkm}) can be summarized as follows:
\begin{equation} 
\frac{T_{k,m}^{k',m'}(a,s+1,u) \, T_{k+1,m+1}^{k',m'}(a,s,u+1) \!- \! T_{k,m}^{k',m'}(a,s,u+1) \,
T_{k+1,m+1}^{k',m'}(a,s+1,u)}{T_{k+1,m}^{k',m'}(a,s,u+1) \, T_{k,m+1}^{k',m'}(a,s+1,u)}\!=\!1, 
\label{TTs}
\end{equation}
\begin{equation} 
\frac{T_{k,m}^{k',m'}(a-1,s,u) \, T_{k+1,m+1}^{k',m'}(a,s,u+1) \!- \! T_{k,m}^{k',m'}(a,s,u+1) \,
T_{k+1,m+1}^{k',m'}(a-1,s,u)}{T_{k+1,m}^{k',m'}(a,s,u+1) \, T_{k,m+1}^{k',m'}(a-1,s,u)}\!=\!1. 
\label{TTa}
\end{equation}
We note that as it was done in \cite{KaziT} further bilinear TT-relations can be derived from (\ref{MTFkm}), 
which will not be presented here.

With the help of the method described above very similar TT-relations can be obtained 
from the left B\"acklund transformations (\ref{MTFk'm'}). The two most important of them read as: 
\begin{equation} 
\frac{T_{k,m}^{k',m'}(a,s-1,u) \, T_{k,m}^{k'+1,m'+1}(a,s,u+1) \!- \! T_{k,m}^{k',m'}(a,s,u+1) \,
T_{k,m}^{k'+1,m'+1}(a,s-1,u)}{T_{k,m}^{k'+1,m'}(a,s,u+1) \, T_{k,m}^{k',m'+1}(a,s-1,u)}\!=\!1, 
\label{TTsv}
\end{equation}
\begin{equation} 
\frac{T_{k,m}^{k',m'}(a-1,s,u) \, T_{k,m}^{k'+1,m'+1}(a,s,u+1) \!- \! T_{k,m}^{k',m'}(a,s,u+1) \,
T_{k,m}^{k'+1,m'+1}(a-1,s,u)}{T_{k,m}^{k'+1,m'}(a,s,u+1) \, T_{k,m}^{k',m'+1}(a-1,s,u)}\!=\!1. 
\label{TTav}
\end{equation}
The so-called QQ-relations are obtained by restricting TT-relations (\ref{TTs}-\ref{TTav}) to the 
boundary of the T-shaped fat hook. The most important ones can be derived by taking (\ref{TTs}) and
(\ref{TTsv}) along the line $(0,s)$. The QQ-relations for the left and right moving functions 
($Q^{k',m'}$ and $\tilde{Q}_{k,m}$) of the lower boundary line of the T-shaped fat-hook respectively  
admit formally the same QQ-relations as the those of the $gl(K|M)$ super spin chains \cite{tJ,KaziT}:
\begin{equation}
\tilde{Q}_{k,m}(u)\,\tilde{Q}_{k+1,m+1}(u+2)-\tilde{Q}_{k,m}(u+2) \, \tilde{Q}_{k+1,m+1}(u)= 
\tilde{Q}_{k,m+1}(u) \, \tilde{Q}_{k+1,m}(u+2),
\label{QQtilde}
\end{equation}
\begin{equation}
Q^{k',m'}(u)\,Q^{k'+1,m'+1}(u+2)-Q^{k',m'}(u+2) \, Q^{k'+1,m'+1}(u)= 
Q^{k',m'+1}(u) \, Q^{k'+1,m'}(u+2). 
  \label{QQv}
\end{equation}
Another QQ-relation can be obtained by taking (\ref{TTsv}) along the line $(k,s)$ for $s\geq m$: 
\begin{equation}
{\cal \tilde{Q}}_k^{k',m'}(u)\,{\cal \tilde{Q}}_k^{k'+1,m'+1}(u+2)-{\cal \tilde{Q}}_k^{k',m'}(u+2) \, {\cal %%@
\tilde{Q}}_k^{k'+1,m'+1}(u)= 
{\cal \tilde{Q}}_k^{k',m'+1}(u) \, {\cal \tilde{Q}}_k^{k'+1,m'}(u+2). 
  \label{QQQ}
\end{equation}
 Finally restricting equations (\ref{TTs}) to the line $(k',s)$ for $s\leq -m'$ one ends up with an additional
type of QQ-relation:
\begin{equation}
{\cal Q}^{k'}_{k,m}(u)\,{\cal Q}^{k'}_{k+1,m+1}(u+2)-{\cal Q}^{k'}_{k,m}(u+2) \, {\cal Q}^{k'}_{k+1,m+1}(u)= 
{\cal Q}^{k'}_{k,m+1}(u) \, {\cal Q}^{k'}_{k+1,m}(u+2). 
  \label{QQB}
\end{equation}
In case of $gl(K|M)$ super spin chains \cite{KaziT} QQ-relations such as (\ref{QQtilde}-\ref{QQB}) were used
to derive Bethe equations for the spin chain. This was possible because in case of spin chains, 
based on earlier experience with algebraic Bethe ansatz, it is assumed that the Q-functions are polynomial
functions of $u$, and then QQ-relations allow one to derive a closed set of equations for their zeroes. 
The zeroes of Q-functions are the Bethe roots which determine uniquely the Q-functions. Then the T-functions
can be expressed in terms of the Q-functions, which provide the solution of the Hirota equations.

In our case, for the T-shaped fat-hook, we do not use QQ-relations to get Bethe equations, because we do not
assume anything about the functional form of the Q-functions. We consider them as general functions as %%@
possible,
which means that they can be arbitrary functions modulo equations which connect them (for example %%@
QQ-relations). 
Our purpose is only to express T-functions in terms of Q-functions, which 
 is presented in the next section.
In this way we can parameterize the infinitely
many T-functions in terms of a few arbitrarily chosen functions. 

In view of future applications these functions have to be dressed by some analytic properties, 
so that one can derive a set of nonlinear integral equations (NLIE\footnote{The nonlinear integral
equation technique (NLIE) proved to be the most efficient tool for describing the finite size effects
in spin chains and relativistic quantum field theories. So, this is the expectation for the AdS/CFT as well.}) 
governing the exact planar spectrum of AdS/CFT.

\section{Integration of the Hirota equations}

In this section following the lines of \cite{KaziT}
 we develop a general algorithm  to solve the Hirota equations
 expressing the functions $T_{k,m}^{k',m'}(a,s,u)$ in terms of boundary functions 
$Q^{k',m'}(u)$ and $\tilde{Q}_{k,m}(u)$. 

The starting point is an alternative representation of the first and second rows of equations (\ref{MTFkm})
and (\ref{MTFk'm'}). 
The equations coming from (\ref{MTFkm}) are formulated by shift operators, which change only the lower indices 
of
$T_{k,m}^{k',m'}(a,s,u)$:    
\begin{equation} \label{TH1}
T_{k-1,m}^{k',m'}(a,s,u)=\hat{H}_{k-,m}^{k',m'}(a-1,s,u) \, T_{k,m}^{k',m'}(a,s,u),
\end{equation}
\begin{equation} \label{TH2}
T_{k,m+1}^{k',m'}(a,s,u)=\hat{H}_{k,m+}^{k',m'}(a-1,s,u) \, T_{k,m}^{k',m'}(a,s,u),
\end{equation}
\begin{equation} \label{TH3}
T_{k+1,m}^{k',m'}(a,s,u)=\hat{H}_{k+,m}^{k',m'}(a,s-1,u) \, T_{k,m}^{k',m'}(a,s,u),
\end{equation}
\begin{equation} \label{TH4}
T_{k,m-1}^{k',m'}(a,s,u)=\hat{H}_{k,m-}^{k',m'}(a,s-1,u) \, T_{k,m}^{k',m'}(a,s,u),
\end{equation}
where the "right shift operators"\footnote{We call these shift operators "right shift operators",
because they change the indices $k$ and $m$, which are the coordinates of the corner point
on the right hand side of the T-shaped fat-hook. The origin of the term "left shift operator"
 is analogous.} take the form:
\begin{equation} \label{Hk-}
\hat{H}_{k-,m}^{k',m'}(a,s,u)=\frac{T_{k-1,m}^{k',m'}(a,s,u+1)}{T_{k,m}^{k',m'}(a,s,u+1)}-
\frac{T_{k-1,m}^{k',m'}(a,s+1,u)}{T_{k,m}^{k',m'}(a,s,u+1)} \, e^{\partial_u-\partial_s},
\end{equation}
\begin{equation} \label{Hm+}
\hat{H}_{k,m+}^{k',m'}(a,s,u)=\frac{T_{k,m+1}^{k',m'}(a,s,u+1)}{T_{k,m}^{k',m'}(a,s,u+1)}-
\frac{T_{k,m+1}^{k',m'}(a,s+1,u)}{T_{k,m}^{k',m'}(a,s,u+1)} \, e^{\partial_u-\partial_s},
\end{equation}
\begin{equation} \label{Hk+}
\hat{H}_{k+,m}^{k',m'}(a,s,u)=\frac{T_{k+1,m}^{k',m'}(a,s,u-1)}{T_{k,m}^{k',m'}(a,s,u-1)}+
\frac{T_{k+1,m}^{k',m'}(a+1,s,u)}{T_{k,m}^{k',m'}(a,s,u-1)} \, e^{-\partial_u-\partial_a},
\end{equation}
\begin{equation} \label{Hm-}
\hat{H}_{k,m-}^{k',m'}(a,s,u)=\frac{T_{k,m-1}^{k',m'}(a,s,u-1)}{T_{k,m}^{k',m'}(a,s,u-1)}+
\frac{T_{k,m-1}^{k',m'}(a+1,s,u)}{T_{k,m}^{k',m'}(a,s,u-1)} \, e^{-\partial_u-\partial_a}.
\end{equation}
The equations coming from (\ref{MTFk'm'}) are formulated by shift operators shifting only 
the upper indices of $T_{k,m}^{k',m'}(a,s,u)$:
\begin{equation} \label{TG1}
T_{k,m}^{k'-1,m'}(a,s,u)=\hat{G}_{k,m}^{k'-,m'}(a-1,s,u) \, T_{k,m}^{k',m'}(a,s,u),
\end{equation}
\begin{equation} \label{TG2}
T_{k,m}^{k',m'+1}(a,s,u)=\hat{G}_{k,m}^{k',m'+}(a-1,s,u) \, T_{k,m}^{k',m'}(a,s,u),
\end{equation}
\begin{equation} \label{TG3}
T_{k,m}^{k'+1,m'}(a,s,u)=\hat{G}_{k,m}^{k'+,m'}(a,s+1,u) \, T_{k,m}^{k',m'}(a,s,u),
\end{equation}
\begin{equation} \label{TG4}
T_{k,m}^{k',m'-1}(a,s,u)=\hat{G}_{k,m}^{k',m'-}(a,s+1,u) \, T_{k,m}^{k',m'}(a,s,u),
\end{equation}
where the "left shift operators" take the form:
\begin{equation} \label{Gk'-}
\hat{G}_{k,m}^{k'-,m'}(a,s,u)=\frac{T_{k,m}^{k'-1,m'}(a,s,u+1)}{T_{k,m}^{k',m'}(a,s,u+1)}-
\frac{T_{k,m}^{k'-1,m'}(a,s-1,u)}{T_{k,m}^{k',m'}(a,s,u+1)} \, e^{\partial_u+\partial_s},
\end{equation}
\begin{equation} \label{Gm'+}
\hat{G}_{k,m}^{k',m'+}(a,s,u)=\frac{T_{k,m}^{k',m'+1}(a,s,u+1)}{T_{k,m}^{k',m'}(a,s,u+1)}-
\frac{T_{k,m}^{k',m'+1}(a,s-1,u)}{T_{k,m}^{k',m'}(a,s,u+1)} \, e^{\partial_u+\partial_s},
\end{equation}
\begin{equation} \label{Gk'+}
\hat{G}_{k,m}^{k'+,m'}(a,s,u)=\frac{T_{k,m}^{k'+1,m'}(a,s,u-1)}{T_{k,m}^{k',m'}(a,s,u-1)}+
\frac{T_{k,m}^{k'+1,m'}(a+1,s,u)}{T_{k,m}^{k',m'}(a,s,u-1)} \, e^{-\partial_u-\partial_a},
\end{equation}
\begin{equation} \label{Gm'-}
\hat{G}_{k,m}^{k',m'-}(a,s,u)=\frac{T_{k,m}^{k',m'-1}(a,s,u-1)}{T_{k,m}^{k',m'}(a,s,u-1)}+
\frac{T_{k,m}^{k',m'-1}(a+1,s,u)}{T_{k,m}^{k',m'}(a,s,u-1)} \, e^{-\partial_u-\partial_a}.
\end{equation}
 Starting from $T_{k,m}^{k',m'}$, with the help of shift operators, the
$T_{k\mp1,m\pm1}^{k',m'}$ and $T_{k,m}^{k'\mp1,m'\pm1}$ functions can be obtained in two different
ways, implying "weak" zero curvature conditions for the shift operators:
\begin{equation}  \label{wzccH}
T_{k\mp1,m\pm1}^{k',m'}=\hat{H}_{k\mp1,m\pm}^{k',m'} \, \hat{H}_{k\mp,m}^{k',m'} \,T_{k,m}^{k',m'}=
\hat{H}_{k\mp,m\pm1}^{k',m'} \, \hat{H}_{k,m\pm}^{k',m'} \,T_{k,m}^{k',m'},
\end{equation}
\begin{equation}  \label{wzccG}
T_{k,m}^{k'\mp1,m'\pm1}=\hat{G}_{k,m}^{k'\mp1,m'\pm} \, \hat{G}_{k,m}^{k'\mp,m'} \,T_{k,m}^{k',m'}=
\hat{G}_{k,m}^{k'\mp,m'\pm1} \, \hat{G}_{k,m}^{k',m'\pm} \,T_{k,m}^{k',m'},
\end{equation}
where for short we skipped the arguments $(a-1,s,u)$ for the shift operators and $(a,s,u)$
for the T-functions. Equations (\ref{wzccH}) and (\ref{wzccG}) are termed "weak" zero curvature conditions,
because they are satisfied only when the shift operators act on the functions $T_{k,m}^{k',m'}$. 
Nevertheless, exploiting the TT-relations (\ref{TTs}-\ref{TTav}),
it can be shown that the "weak" zero curvature conditions imply the strong operator form as well: 
\begin{equation}  \label{szccH}
\hat{H}_{k\mp1,m\pm}^{k',m'}(a,s,u) \, \hat{H}_{k\mp,m}^{k',m'} (a,s,u)=
\hat{H}_{k\mp,m\pm1}^{k',m'} (a,s,u)\, \hat{H}_{k,m\pm}^{k',m'}(a,s,u),
\end{equation}
\begin{equation}  \label{szccG}
\hat{G}_{k,m}^{k'\mp1,m'\pm}(a,s,u) \, \hat{G}_{k,m}^{k'\mp,m'}(a,s,u) =
\hat{G}_{k,m}^{k'\mp,m'\pm1}(a,s,u) \, \hat{G}_{k,m}^{k',m'\pm}(a,s,u).
\end{equation}
Having the above relations at hand we express the T-functions in terms of the $Q^{k',m'}$
and $\tilde{Q}_{k,m}$ boundary functions. To avoid problems coming from
division by zero in the shift operators, we determine the T-functions along the infinite line
$(1,s)$ and then T-functions for other values of $a$ and $s$ can be determined by the application
of the Bazhanov-Reshetikhin formula \cite{BR,Tsuboi}: 
\begin{equation} \label{BR}
\tilde{T}_{k,m}^{k',m'}(a,s,u)=
\displaystyle{\det_{1\leq i,j \leq a}} \, \tilde{T}_{k,m}^{k',m'}(1,s+i-j,u+a+1-i-j),
\end{equation} 
where
\begin{equation}
\tilde{T}_{k,m}^{k',m'}(a,s,u)=\frac{T_{k,m}^{k',m'}(a,s,u)}{Q^{k',m'}(u+a+s) \,\tilde{Q}_{k,m}(u-a-s)}.
\end{equation}
To express $T_{k,m}^{k',m'}(1,s,u)$ in terms of $Q^{k',m'}$ and $\tilde{Q}_{k,m}$, the
equations (\ref{Hk-}), (\ref{Hm+}) and (\ref{Gk'-}), (\ref{Gm'+}) have to be taken at $a=1$.
In this case the corresponding shift operators depend only on the boundary functions 
$Q^{k',m'}$ and $\tilde{Q}_{k,m}$. Moreover it turns out that for $a=0$ the right shift operators
depend only on the lower indices $k$ and $m$, while the left shift operators depend only on  the
upper indices $k'$ and $m'$. This is why we delete the irrelevant pair of indices from them.
The relevant shift operators are of the form:
\begin{equation} \label{H0sk-}
\hat{H}_{k-,m}(0,s,u)=\frac{\tilde{Q}_{k-1,m}(u-s+1)}{\tilde{Q}_{k,m}(u-s+1)}
-\frac{\tilde{Q}_{k-1,m}(u-s-1)}{\tilde{Q}_{k,m}(u-s+1)} \, e^{\partial_u-\partial_s},
\end{equation}
\begin{equation} \label{H0sm+}
\hat{H}_{k,m+}(0,s,u)=\frac{\tilde{Q}_{k,m+1}(u-s+1)}{\tilde{Q}_{k,m}(u-s+1)}
-\frac{\tilde{Q}_{k,m+1}(u-s-1)}{\tilde{Q}_{k,m}(u-s+1)} \, e^{\partial_u-\partial_s},
\end{equation}
\begin{equation} \label{G0sk'-}
\hat{G}^{k'-,m'}(0,s,u)=\frac{Q^{k'-1,m'}(u+s+1)}{Q^{k',m'}(u+s+1)}
-\frac{Q^{k'-1,m'}(u+s-1)}{Q^{k',m'}(u+s+1)} \, e^{\partial_u+\partial_s},
\end{equation}
\begin{equation} \label{G0sm'+}
\hat{G}^{k',m'+}(0,s,u)=\frac{Q^{k',m'+1}(u+s+1)}{Q^{k',m'}(u+s+1)}
-\frac{Q^{k',m'+1}(u+s-1)}{Q^{k',m'}(u+s+1)} \, e^{\partial_u+\partial_s}.
\end{equation}
It is important to recognize that the left shift operators depend only on $u+s$ and the
right ones on $u-s$, which implies that along this special line the left and right shift operators
commute for any indices:
\begin{equation} \label{GH}
\hat{G}^{\alpha}(0,s,u) \,\hat{H}_{\beta}(0,s,u)=\hat{H}_{\beta}(0,s,u) \, \hat{G}^{\alpha}(0,s,u), 
\end{equation}
where $\alpha$ denotes either the pair of indices "$k',m'+$" or "$k'-,m'$", while similarly
 $\beta$ is either for the pair of indices "$k,m+$" or "$k-,m$". 

Relations (\ref{GH}) ensure that during the undressing procedure the size of the system on the left and
right hand sides can be decreased independently. Moreover equations (\ref{szccH}) and (\ref{szccG}) 
ensure that the different orders of decreasing the values of $k,m$ and $k',m'$ respectively lead to the same
final result. 

Thus to determine an undressing path, one can define two arbitrary oriented zigzag paths; 
one going from $(0,0)$ to $(K',-M')$ denoted
by $\gamma^{K',M'}$, and the other $\gamma_{K,M}$ goes from $(0,0)$ to $(K,M)$.
 These two paths are characterized by their oriented edges, in such a way that to each edge a pair of
vectors $({\bf x,n})$ is assigned, where ${\bf x}$ denotes the starting point of the oriented edge 
on the $(a,s)$ lattice, while  ${\bf n}$ is a unit vector pointing from ${\bf x}$ to the endpoint
of the edge. For the path $\gamma^{K',M'}$ the edge characterizing pairs of vectors are denoted by
$({\bf x_L,n_L})$, while for $\gamma_{K,M}$ they are denoted by $({\bf x_R,n_R})$.
The unit vectors ${\bf n_L}$ can take values $(1,0)$ or $(0,-1)$, while the possible choices for
${\bf n_R}$ are $(1,0)$ or $(0,1)$.

Using the previous characterization, inverse undressing operators\footnote{These operators are called "inverse 
undressing operators", because they realize the inverse of the undressing procedure, which means that
their successive application to the trivial element $T_{0,0}^{0,0}(a,s,u)$ of the hierarchy of T-functions 
 provides the solution to the original problem.}
 can be assigned to the edges of the
paths. They are as follows:
\begin{equation} \label{VR}
\hat{V}^{R}_{({\bf x_R,n_R})}(s,u)=\left\{
\begin{array}{cc}
\hat{H}_{k,m+}(0,s,u), \qquad {\bf x_R}=(k,m), \quad {\bf n_R}=(0,1) \\
\hat{H}_{(k+1)-,m}^{-1}(0,s,u), \qquad {\bf x_R}=(k,m), \quad {\bf n_R}=(1,0),
\end{array}
 \right.
\end{equation}
\begin{equation} \label{VL}
\hat{V}^{L}_{({\bf x_L,n_L})}(s,u)=\left\{
\begin{array}{cc}
\hat{G}^{k',m'+}(0,s,u), \qquad {\bf x_L}=(k',-m'), \quad {\bf n_L}=(0,-1) \\
(\hat{G}^{(k'+1)-,m'}(0,s,u))^{-1}, \qquad {\bf x_L}=(k',-m'), \quad {\bf n_L}=(1,0),
\end{array}
 \right.
\end{equation}
where $\hat{V}^{L}_{({\bf x_L,n_L})}(s,u)$ is assigned to the edge $({\bf x_L,n_L})$ of
$\gamma^{K',M'}$, while $\hat{V}^{R}_{({\bf x_R,n_R})}(s,u)$ is assigned to the edge  
$({\bf x_R,n_R})$ of $\gamma_{K,M}$.
In equations (\ref{VR}) and (\ref{VL}) the inverse operators of $\hat{H}_{(k+1)-,m}(0,s,u)$ and
$\hat{G}^{(k'+1)-,m'}(0,s,u)$ appear. It can be shown \cite{KaziT} that these operators have nontrivial
kernels. Namely, the kernel of $\hat{H}_{k-,m}(0,s,u)$
consists of functions of the form $\tilde{Q}_{k-1,m}(u-s-1) \, \tilde{f}(u+s)$ and the kernel of
$\hat{G}^{k'-,m'}(0,s,u)$ consists of functions of the form $Q^{k'-1,m}(u+s-1) \, f(u-s)$, 
with $f(u)$ and $\tilde{f}(u)$  being  arbitrary functions. In formulae (\ref{VR}) and (\ref{VL})
the inverses of $\hat{H}_{k-,m}(0,s,u)$ and $\hat{G}^{k'-,m'}(0,s,u)$ are meant modulo these kernels.

Following from equations (\ref{TH1}),(\ref{TH2}) (\ref{TG1}),(\ref{TG2}) taken at $a=1$, the T-functions %%@
$T_{K,M}^{K',M'}(1,s,u)$ can be obtained as ordered products of operators 
(\ref{VR}) and (\ref{VL}) acting on $T_{0,0}^{0,0}(1,s,u)$:
\begin{equation} \label{TKMK'M'}
T_{K,M}^{K',M'}(1,s,u)=\hat{\cal V}_L(s,u) \, \hat{\cal V}_R(s,u) \, T_{0,0}^{0,0}(1,s,u),
\end{equation}
where the operators $\hat{\cal V}_L(s,u)$ and $\hat{\cal V}_R(s,u)$ commute due to (\ref{GH}), and they
correspond to paths $\gamma^{K',M'}$ and $\gamma_{K,M}$ respectively. They read as
\begin{equation} \label{calVL}
\hat{\cal V}_L(s,u)=\prod\limits_{({\bf x_L,n_L})\in \gamma^{K',M'}}^{\rightarrow} \,
 \hat{V}^{L}_{({\bf x_L,n_L})}(s,u),
\end{equation}
\begin{equation} \label{calVR}
\hat{\cal V}_R(s,u)=\prod\limits_{({\bf x_R,n_R})\in \gamma_{K,M}}^{\rightarrow} \,
 \hat{V}^{R}_{({\bf x_R,n_R})}(s,u),
\end{equation}
where the arrow means that ordered product of the operators has to be taken, in such a way
that operators on the right hand side of (\ref{calVL}) and (\ref{calVR}) have to be ordered 
from the last edge of the corresponding path (ending at the origin). I.e. reading the product
of operators on the right hand side of (\ref{calVL}) and (\ref{calVR}) from the left to the right,
the first operator corresponds to the last edge of the oriented paths 
$\gamma^{K',M'}$ and $\gamma_{K,M}$ respectively.
%%%%%%%%%%%%%%%%%%%%%%%%%%%%%%%%%%%%%%%
\begin{figure}[htb]
\begin{flushleft}
%\vskip 10mm
\hskip 15mm
\leavevmode
\epsfxsize=120mm
\epsfbox{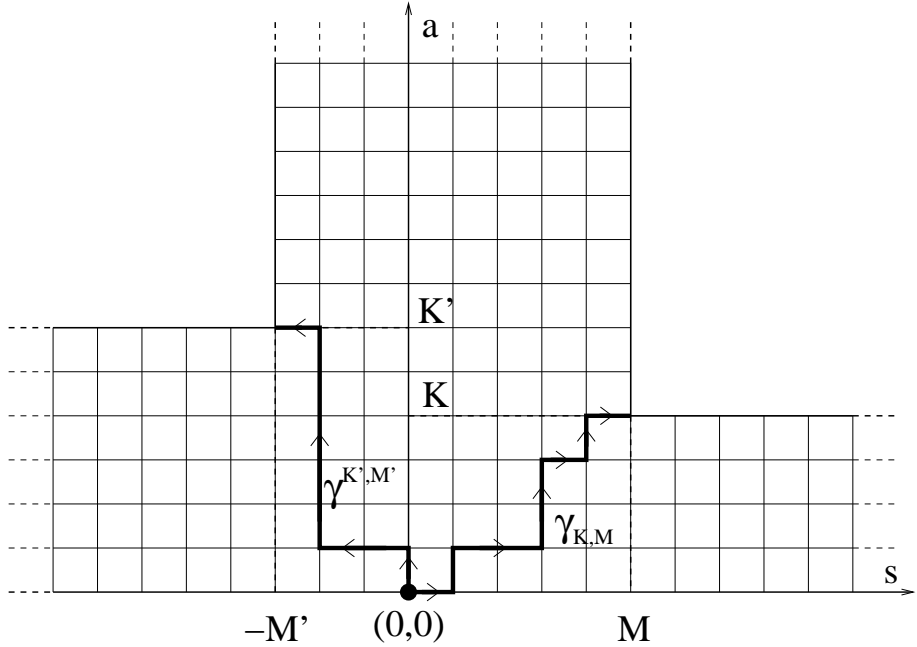}
%\vskip 10mm
\end{flushleft}
\caption{{\footnotesize
Undressing paths for a T-shaped fat-hook of type $(K',M')\odot (K,M)$. 
}}
\label{4}
\end{figure}
%%%%%%%%%%%%%%%%%%%%%%%%%%%%%%%%%%%%%%%%% 
 The T-function $T_{0,0}^{0,0}(1,s,u)$ corresponds to a trivial domain, thus it can be trivially
expressed by the boundary functions (\ref{boundarykm}):
\begin{equation} \label{T0000}
T_{0,0}^{0,0}(1,s,u)=Q^{0,0}(u+1)\,\tilde{Q}_{0,0}(u-1) \, \delta_{s,0}=Q^{0,0}(u-1)\,\tilde{Q}_{0,0}(u+1) \, \delta_{s,0}.
\end{equation}

The formulae (\ref{TKMK'M'})-(\ref{T0000}) provide the solution of the Hirota equations defined on a T-shaped
fat-hook of type $(K',M') \odot (K,M)$, in such a way that the T-functions are expressed in terms of
the boundary functions (Q-functions) of the hierarchy of Hirota equations defined by our chain of
B\"acklund transformations.

 The application of formula (\ref{TKMK'M'}) to the case of AdS/CFT (i.e. $K=M=K'=M'=2$)
will be presented in the next section. One remark concerning equation (\ref{TKMK'M'}) is in order.
It is not true that the functions $T_{K,M}^{K',M'}(1,s,u)$ are products of eigenvalues of 
a $gl(K|M)$ and a $gl(K'|M')$ transfer matrices, but rather they can be obtained by acting a $gl(K|M)$ type
($\hat{\cal V}_R$) and a $gl(K'|M')$ type ($\hat{\cal V}_L$) generating series on $T_{0,0}^{0,0}(1,s,u)$.  
This fact accounts for a possible $GL(K|M)\otimes GL(K'|M')$ symmetry\footnote{ In this remark the symmetry 
means that this T-system possibly correspond to the
TBA equations of an appropriate $GL(K|M)\otimes GL(K'|M')$ invariant scattering theory.} 
 behind the T-system defined on a T-shaped fat-hook of type $(K',M')\odot(K,M)$.

The section is closed with a remark concerning the arbitraryness of the undressing procedure.
In the undressing procedure in addition to the arbitrary choice of the paths $\gamma^{K',M'}$
and $\gamma_{K,M}$ there is an additional arbitraryness.
 Namely 
the origin of our state of reference on the $(a,s)$ lattice can be shifted freely within the interval
$[-M',M]$. This means, that for example the T-shaped fat-hook with corner point coordinates 
$(K',-M')$ and $(K,M)$ is equivalent to T-shaped fat-hooks with corner point coordinates 
$(K',-M'+m)$ and $(K,M+m)$, where $-M\leq m \leq M'$. (We restricted the possible values for $m$, 
which is the origin shifting parameter, so that the origin will remain between the "s-coordinates" 
of the corner points.)
This means that the number of zigzag paths defining the undressing procedure is:
\begin{equation} \label{szam}
N_{zigzag}=\sum\limits_{j=0}^{M+M'} 
\left(
\begin{array}{cc} 
K'+M+M'-j \\
K'
\end{array}
\right)
\left(
\begin{array}{cc} 
K+j \\
K
\end{array}
\right),
\end{equation}
 where $\left(
\begin{array}{cc} 
n \\
m
\end{array}
\right)$
stands for the binomial coefficient.  In case of AdS/CFT $(K=K'=M=M'=2)$ this counting gives 
 $N_{zigzag}^{AdS}=\left(
\begin{array}{cc} 
9 \\
5
\end{array}
\right)$ for the total number of possible undressing paths.

However it has to be noted that there may be such distinguished axes on the $(a,s)$
lattice, which distinguish certain choices for the origin.
 In case of AdS/CFT the analytic properties of Y-functions distinguish the vertical symmetry axis 
of the T-shaped fat-hook of type $(2,2) \odot (2,2)$. This is why we solve the AdS/CFT case from a state
of reference, where the corner point coordinates of the T-shaped fat-hook are $(2,-2)$ and $(2,2)$ on the
$(a,s)$ lattice. In this case the distinguished vertical symmetry axis of the T-shaped fat-hook is a vertical
line starting from the origin. This choice for the origin is made, so that none of the zigzag paths
 will cross the distinguished axis.

\section{The AdS/CFT case}

In this section the method explained in the previous sections is applied to the case of AdS/CFT.
Based on TBA calculations \cite{KaziY1,KaziY2} the T-system  (\ref{T}) of AdS/CFT is defined on a T-shaped
fat-hook of type $(2,2)\odot(2,2)$ as a consequence of the $SU(2|2)\otimes SU(2|2)$ symmetry of
the problem. Here we make  special choices for the undressing paths $\gamma^{K',M'}$ and $\gamma_{K,M}$
drawn in figure 7., and then we express the T-functions of the problem along the line $(1,s)$ in terms of
the boundary functions $Q^{k',m'}$ and $\tilde{Q}_{k,m}$. 
%%%%%%%%%%%%%%%%%%%%%%%%%%%%%%%%%%%%%%%
\begin{figure}[htb]
\begin{flushleft}
%\vskip 10mm
\hskip 15mm
\leavevmode
\epsfxsize=100mm
\epsfbox{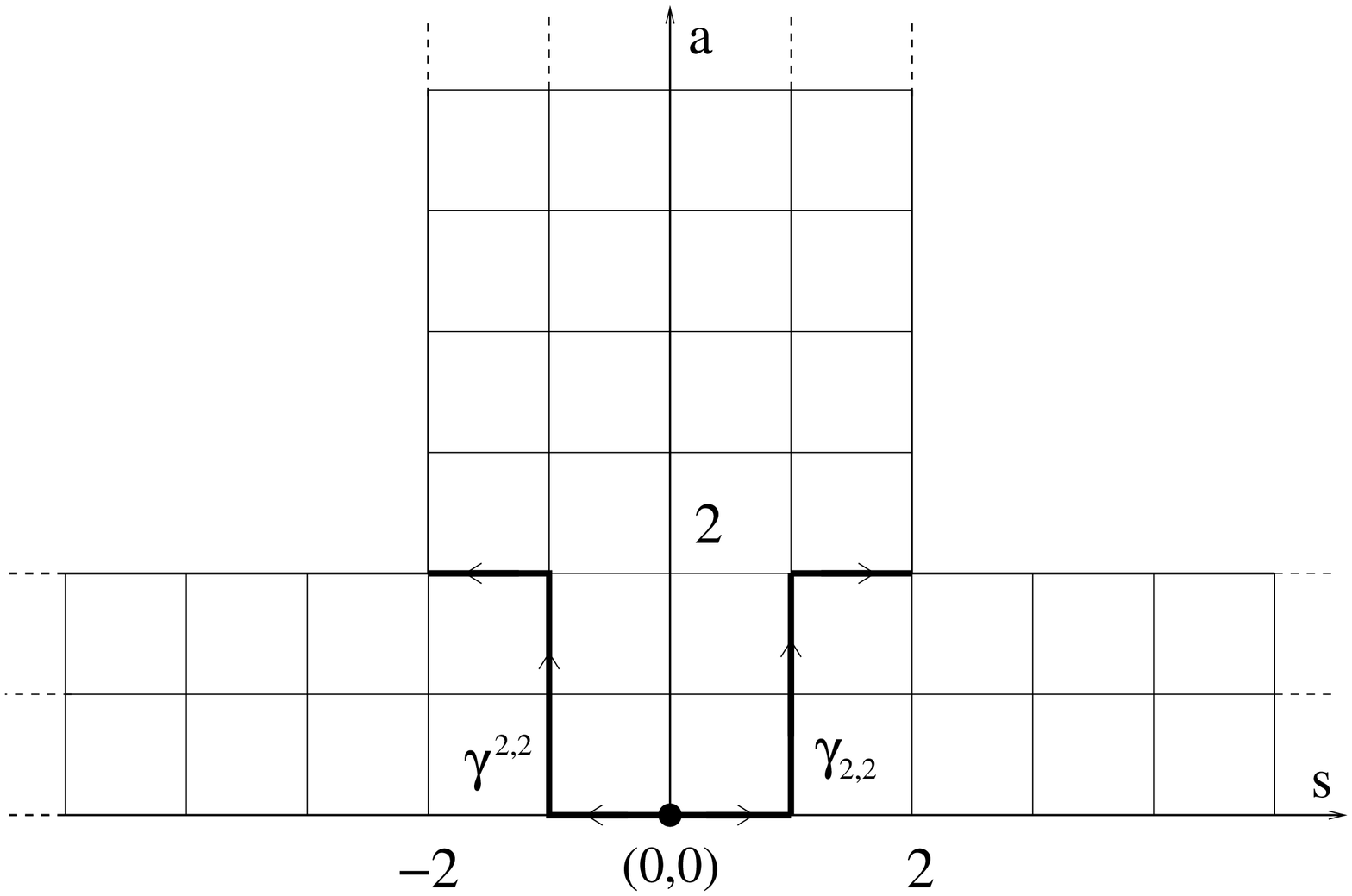}
%\vskip 10mm
\end{flushleft}
\caption{{\footnotesize
A possible undressing path for AdS/CFT. 
}}
\label{7}
\end{figure}
%%%%%%%%%%%%%%%%%%%%%%%%%%%%%%%%%%%%%%%%%
The generating series corresponding to the path  $\gamma_{K,M}$ takes the form:
\begin{equation} \label{VR22}
\hat{\cal V}_R(s,u)=\hat{H}_{2,1+}(0,s,u) \,\hat{H}_{2-,1}^{-1}(0,s,u) \,\hat{H}_{1-,1}^{-1}(0,s,u) \,
\hat{H}_{0,0+}(0,s,u)  
\end{equation}
while the generating series for $\gamma^{K',M'}$ reads as:
\begin{equation} \label{VL22}
\hat{\cal V}_L(s,u)=\hat{G}^{2,1+}(0,s,u) \,(\hat{G}^{2-,1}(0,s,u))^{-1} \,(\hat{G}^{1-,1}(0,s,u))^{-1} \,
\hat{G}^{0,0+}(0,s,u).  
\end{equation}
Expanding the inverse operators  with respect to shift operators in $u$ and $s$ 
and collecting them to the right hand side of the expressions, one obtains the following result 
for the generating series $\hat{\cal V}_L(s,u)$  and $\hat{\cal V}_R(s,u)$:
\begin{equation} \label{VLR}
\hat{\cal V}_{\Omega}(s,u)=\sum\limits_{k=0}^{\infty} F_{\Omega}^{(k)}(u+\sigma \, s) \, 
e^{k(\partial_u+\sigma \, \partial_s )}, \qquad
\sigma=\left\{\begin{array}{cc}
+1  & \Omega=L \\
-1  & \Omega=R
\end{array}
\right.
\end{equation}
where $\Omega$ stands for $L$ or $R$ and $\sigma$ can take values $\pm 1$ depending on the 
value of lower index ($L$ or $R$) of the generating series $\hat{\cal V}_{\Omega}(s,u)$.
The coefficients in (\ref{VLR}) are of the form:
\begin{eqnarray}
F_{\Omega}^{(k)}(u)=E_{\Omega}(u) \, \sum\limits_{j=0}^k \, A_{\Omega}(u+2(k-j)) \,
 B_{\Omega}(u+2k)-
\nonumber \\
\Theta(k-1) \, E_{\Omega}(u) \, \sum\limits_{j=0}^{k-1} \, A_{\Omega}(u+2(k-j-1)) \, C_{\Omega}(u+2(k-1))- 
\nonumber \\
\Theta(k-1) \, F_{\Omega}(u) \, \sum\limits_{j=0}^{k-1} \, A_{\Omega}(u+2(k-j)) \, 
B_{\Omega}(u+2k)+
\nonumber \\
\Theta(k-2) \, F_{\Omega}(u) \, \sum\limits_{j=0}^{k-2} \, A_{\Omega}(u+2(k-j-1)) \, C_{\Omega}(u+2(k-1)),
\end{eqnarray}
where the $A_{\Omega},B_{\Omega},C_{\Omega},E_{\Omega},F_{\Omega}$ coefficient functions can be directly
expressed by the boundary functions of the problem:
\begin{equation}
A_{\Omega}(u)=\frac{ {\bf Q}_{\Omega}^{2,1}(u+1) \,{\bf Q}_{\Omega}^{0,1}(u-1) }{{\bf Q}_{\Omega}^{1,1}(u+1) %%@
\,{\bf Q}_{\Omega}^{1,1}(u-1)}, \qquad 
B_{\Omega}(u)=\frac{ {\bf Q}_{\Omega}^{1,1}(u+1) }{{\bf Q}_{\Omega}^{0,1}(u-1) \,{\bf Q}_{\Omega}^{0,0}(u+1)},
\end{equation}
\begin{equation}
 C_{\Omega}(u)=\frac{ {\bf Q}_{\Omega}^{1,1}(u+1) }{{\bf Q}_{\Omega}^{0,1}(u+1) \,{\bf Q}_{\Omega}^{0,0}(u+1)},
\qquad
E_{\Omega}(u)=\frac{ {\bf Q}_{\Omega}^{2,2}(u+1) \, {\bf Q}_{\Omega}^{1,1}(u-1) }{{\bf Q}_{\Omega}^{2,1}(u+1)}
\end{equation}
\begin{equation}
F_{\Omega}(u)=\frac{ {\bf Q}_{\Omega}^{2,2}(u+1) \, {\bf Q}_{\Omega}^{1,1}(u+1) }{{\bf Q}_{\Omega}^{2,1}(u+1)},
\qquad \Omega=L,R
\end{equation}
where we introduced the notation:
\begin{equation}
{\bf Q}_{\Omega}^{k,m}(u)=\left\{ 
\begin{array}{cc}
\tilde{Q}_{k,m}(u) & \mbox{in case of } \quad \Omega=R \\ 
Q^{k,m}(u) & \mbox{in case of} \quad \Omega=L.
\end{array}
\right.
\end{equation}
Applying the formulae (\ref{TKMK'M'}) and (\ref{T0000}) and putting everything together one obtains 
series expressions for the T-functions along the line $(1,s)$:
\begin{equation} \label{T2222}
T_{2,2}^{2,2}(1,s,u)=\sum\limits_{k=\mbox{{\scriptsize max}}(-s,0)}^{\infty} F_L^{(k)}(u+s) \, F_R^{(s+k)}(u-s) 
\,
Q^{0,0}(u+s+2k+1) \, \tilde{Q}_{0,0}(u+s+2k-1).
\end{equation}
It can be seen that the T-functions along the line $(1,s)$ can be given only as an infinite series
of Q-functions. T-functions for other values of $a$ and $s$ can be determined from (\ref{T2222})
by the application of the Bazhanov -Reshetikhin formula (\ref{BR}). As a final result we could
express the infinitely many T-functions in terms of 9 boundary functions of the hierarchy
of Hirota equations of the T-shaped fat-hook of type $(2,2)\odot(2,2)$. 

An important remark concerning (\ref{T2222}) is in order. In (\ref{T2222}) $T_{2,2}^{2,2}(1,s,u)$ is
represented by an infinite series, so the convergence of the series has to be dealt with. 
Obviously for arbitrary 
boundary functions the series (\ref{T2222}) can not be expected to be convergent, but it may be 
convergent (at least on certain intervals in $u$) for certain choices of boundary functions 
being consistent with QQ-relations (\ref{QQtilde}),(\ref{QQv}).
Anyway, it can be seen that the usual polynomial spin-chain ansatz for all the boundary functions
would lead to divergence. This fact is in accordance with the observation of refs. \cite{KaziY1,KaziO4}, which
states that when the the length of the spin chain in AdS/CFT tends to infinity
$T_{a,s \leq 0}(u)$ and $T_{a,s \geq 0}(u)$ cannot be simultaneously finite.

Despite the fact that the formula (\ref{T2222}) is difficult to apply in practise, 
we think that the algorithm for the integration of the Hirota equations described in the previous section
 together with the TT-, TQ-, and QQ-relations
can be used as a tool for the derivation of an NLIE governing the exact finite size
spectrum of anomalous dimensions of ${\cal N}=4$ SYM in the planar limit.

\section{Summary and discussion}

In this paper we dealt with the solution of Hirota equations (\ref{T}) defined on a T-shaped fat-hook
with corner point coordinates $(K,M)$ and $(K',-M')$.
The key ingredient to the solution is the existence of a set of auto-B\"acklund transformations
for the equations, i.e. transformations which send any solution of the equations 
to another solution.
We introduced four B\"acklund transformations. 
They send any solution of 
the Hirota equations defined on the T-shaped fat-hook with corner point coordinates $(K,M)$ and $(K',-M')$
to a solution corresponding to a T-shaped fat hook with different corner point 
coordinates. (The coordinates $K$,$M$, $K'$, $M'$ differ.)
Specifically, the first B\"acklund transformation (BT1) lowers $K$ by $1$, the second one (BT2) lowers
$M$ by $1$, the third one ($\overline{BT1}$) lowers $K'$ by $1$, and finally the fourth one 
($\overline{BT2}$) lowers $M'$ by $1$, leaving the other corner point coordinates intact. 
Thus they define a hierarchy of Hirota equations.
Applying the four B\"acklund transformations successively $K+M+K'+M'$ times, one comes to a collapsed
 domain being the union of two lines. In this way the original problem gets undressed to a trivial one.
In case of AdS/CFT, when $K=M=K'=M'=2$, the different orders, in which one decreases the original T-shaped 
fat-hook
to a trivial domain, correspond to  different choices of the basis of simple roots of the symmetry
group $SU(2|2)\otimes SU(2|2)$.

With the help of the four B\"acklund transformations we derived TT-, TQ-, and QQ-relations for our system and 
inverting the undressing procedure we could express the infinitely many T-functions of the problem
in terms of a few arbitrary functions characterizing the boundary conditions of the hierarchy
 of Hirota equations (Q-functions).

 In \cite{FrolSuz} it has been shown that since the Y-functions have infinitely many cuts on the $u$ plane
  it is 
natural to think that the Y-system is in fact defined on an infinite genus Riemann surface and the Y-system 
equations (\ref{Y}) hold only in the strip 
$-2 g \leq \mbox{Im} u \leq 2 g$ of the $u$ plane\footnote{Adapting the notations of \cite{FrolSuz} to those of %%@
this paper} . In this case the T- and Q-functions have infinitely many 
branch cuts on the $u$ plane and can also be thought to be defined on infinite genus Riemann surfaces. 

The knowledge of the analytic properties of the Y-functions on the infinite genus Riemann surface would allow 
one to reproduce the TBA equations purely from the Y-system equations. Similarly according to our expectations,
 the understanding of the analytic properties of the T- and Q-functions on the infinite genus Riemann surface 
would make it possible to combine the T- and Q-functions of the problem into a new set of functions which have
 appropriate analytic properties in $u$ to serve as unknown functions of a desired NLIE 
governing the spectrum of anomalous  dimensions in planar ${\cal N}=4$ SYM. 

Another application of the results of this paper could be  the construction of
a lattice model or a spin chain, which probably after some limiting procedure would account for
the spectrum of AdS/CFT.

\vspace{1cm}
{\tt Acknowledgments}

\noindent 
This investigation was supported by the Hungarian National Science Fund OTKA (under T049495).

\section*{Appendix A}

In this appendix the relations between boundary functions of (\ref{boundarykm})
dictated by "degenerated" cases are listed (see the end of section 3.). 
The relations are as follows:
\begin{eqnarray} 
{\cal \tilde{Q}}_{0}^{k',m'}(u)&=&Q^{k',m'}(u), \qquad {\cal Q}_{k,m}^0(u)=\tilde{Q}_{k,m}(u), \qquad
 \nonumber \\
 {\cal Q}_{k,m}^k(u)&=&\tilde{Q}_{0,m}(u-2k), \qquad 
{\cal \tilde{Q}}_{k}^{k,m}(u)=Q^{0,m}(u-2k), \nonumber \\
{\cal \tilde{Q}}_k^{k',m}(u) &=& (-1)^{-m(k+k')} \, {\cal Q}_{k,-m}^{k'}(u+2m) \qquad %%@
\tilde{Q}_{0,m}(u)=Q^{0,-m}(u+2m). 
\end{eqnarray}

\section*{Appendix B}

In this appendix we list additional TQ-type identities between the T-functions and the boundary characterizing
functions. These bilinear identities can be derived from the (\ref{MTFkm}) and (\ref{MTFk'm'}) B\"acklund
transformations by taking them at lines lying close to the boundaries.

Taking the second row of (BT1) (\ref{MTFkm}) along the lines $(a,m-1)$ and $(k-1,s)$, one gets the following 
bilinear relations:
 \begin{eqnarray}
T_{k-1,m}^{k',m'}(a,m-1,u) \, {\cal \tilde{Q}}_k^{k',m'}(u+a+m+1)-(-1)^m \, T_{k,m}^{k',m'}(a,m-1,u) \times  \\ 
{\cal \tilde{Q}}_{k-1}^{k',m'}(u+a+m+1) = T_{k,m}^{k',m'}(a+1,m-1,u+1) \,  {\cal %%@
\tilde{Q}}_{k-1}^{k',m'}(u+a+m-1), \quad a\geq k, \nonumber
\end{eqnarray}
\begin{eqnarray}
T_{k,m}^{k',m'}(k-1,s+1,u+1) \, {\cal \tilde{Q}}_{k-1}^{k',m'}(u+s+k-1)-T_{k,m}^{k',m'}(k-1,s,u) \times  \\ 
{\cal \tilde{Q}}_{k-1}^{k',m'}(u+s+k+1) = T_{k-1,m}^{k',m'}(k-2,s+1,u) \,  {\cal %%@
\tilde{Q}}_{k}^{k',m'}(u+s+k+1), \quad s\geq m. \nonumber
\end{eqnarray}
Taking the first row of (BT2) (\ref{MTFkm}) at $s=m-1$ and at $a=k-1$, one gets the following 
TQ-type equations: 
\begin{eqnarray}
T_{k,m}^{k',m'}(a,m-1,u+1) (-1)^{m-1}  \tilde{Q}_{0,m-1}(u-m-a)-T_{k,m}^{k',m'}(a+1,m-1,u) \times \quad
 \\
\tilde{Q}_{0,m-1}(u-a-m+2) =
T_{k,m-1}^{k',m'}(a+1,m-2,u+1) (-1)^{a-k} \tilde{Q}_{0,m}(u-a-m), \, a\geq k,  \nonumber
\end{eqnarray} 
\begin{eqnarray}
T_{k,m}^{k',m'}(k-1,s,u+1)  \tilde{Q}_{0,m-1}(u-s-k)-T_{k,m-1}^{k',m'}(k-1,s,u+1) \times \quad
 \\
\tilde{Q}_{0,m}(u-s-k) =
T_{k,m}^{k',m'}(k-1,s+1,u) \tilde{Q}_{0,m-1}(u-s-k+2), \quad s\geq m.  \nonumber
\end{eqnarray} 
Taking the second row of (\ref{MTFk'm'}) both for ($\overline{BT1}$) and ($\overline{BT2}$)
along the line $(a,m)$ for $a \geq k$ one gets the following 
bilinear relations:
\begin{eqnarray}
 T_{k,m}^{k',m'}(a,m-1,u+1) {\cal \tilde{Q}}_k^{k'-1,m'}(u+a+m) - T_{k,m}^{k'-1,m'}(a,m-1,u+1) \times \quad \\
  {\cal \tilde{Q}}_{k}^{k',m'}(u+a+m)=(-1)^{m}  \, T_{k,m}^{k'-1,m'}(a-1,m-1,u) \, {\cal %%@
\tilde{Q}}_{k}^{k',m'}(u+a+m+2), \, a\geq k,
\nonumber
\end{eqnarray}
\begin{eqnarray}
 T_{k,m}^{k',m'-1}(a,m-1,u+1) {\cal \tilde{Q}}_k^{k',m'}(u+a+m) -  T_{k,m}^{k',m'}(a,m-1,u+1) \times  \\ 
  {\cal \tilde{Q}}_{k}^{k',m'-1}(u+a+m)=(-1)^{m}  \, T_{k,m}^{k',m'}(a-1,m-1,u) \, {\cal %%@
\tilde{Q}}_{k}^{k',m'-1}(u+a+m+2), \, a\geq k. \nonumber
\end{eqnarray}
Taking the first row of (\ref{MTFk'm'}) both for ($\overline{BT1}$) and ($\overline{BT2}$)
at $a=k-1$, one gets the following equations:
\begin{eqnarray}
 T_{k,m}^{k'-1,m'}(k-1,s,u+1) {\cal \tilde{Q}}_k^{k',m'}(u+s+k) - T_{k,m}^{k',m'}(k-1,s,u+1) \times \quad \\
  {\cal \tilde{Q}}_{k}^{k'-1,m'}(u+s+k)= T_{k,m}^{k'-1,m'}(k-1,s-1,u) \, {\cal \tilde{Q}}_{k}^{k',m'}(u+s+k+2), %%@
\quad s\geq m,
\nonumber
\end{eqnarray}
\begin{eqnarray}
 T_{k,m}^{k',m'}(k-1,s,u+1) {\cal \tilde{Q}}_k^{k',m'-1}(u+s+k) - T_{k,m}^{k',m'-1}(k-1,s,u+1) \times \quad \\
  {\cal \tilde{Q}}_{k}^{k',m'}(u+s+k)= T_{k,m}^{k',m'}(k-1,s-1,u) \, {\cal \tilde{Q}}_{k}^{k',m'-1}(u+s+k+2), %%@
\quad s\geq m.
\nonumber
\end{eqnarray}
Taking the second rows of (\ref{MTFkm}) at $s=-m'$ for both the (BT1) and (BT2) cases, the following 
bilinear equations emerge for $a \geq k'$:
\begin{eqnarray}
T_{k,m}^{k',m'}(a,-m'+1,u+1) \, {\cal Q}^{k'}_{k-1,m}(u+a+m') - T_{k-1,m}^{k',m'}(a,-m'+1,u+1) \times \qquad \\  %%@
{\cal Q}^{k'}_{k,m}(u+a+m')=(-1)^{m'}  \, T_{k-1,m}^{k',m'}(a-1,-m'+1,u) \, {\cal Q}^{k'}_{k,m}(u+a+m'+2), %%@
\quad a\geq k',
\nonumber
\end{eqnarray}
\begin{eqnarray}
T_{k,m-1}^{k',m'}(a,-m'+1,u+1) \, {\cal Q}^{k'}_{k,m}(u+a+m') - T_{k,m}^{k',m'}(a,-m'+1,u+1) \times \qquad \\  %%@
{\cal Q}^{k'}_{k,m-1}(u+a+m')=(-1)^{m'}  \, T_{k,m}^{k',m'}(a-1,-m'+1,u) \, {\cal Q}^{k'}_{k,m-1}(u+a+m'+2), %%@
\quad a\geq k'.
\nonumber
\end{eqnarray}
Taking the first rows of (\ref{MTFkm}) at $a=k'-1$ for both the (BT1) and (BT2) cases, one obtains the 
following TQ-type relations for $s\leq -m'$:
\begin{eqnarray}
T_{k-1,m}^{k',m'}(k'-1,s,u+1) \, {\cal Q}^{k'}_{k,m}(u-s+k') - T_{k,m}^{k',m'}(k'-1,s,u+1) \times \qquad \\  %%@
{\cal Q}^{k'}_{k-1,m}(u-s+k')= T_{k-1,m}^{k',m'}(k'-1,s+1,u) \, {\cal Q}^{k'}_{k,m}(u-s+k'+2), \quad s\leq -m',
\nonumber
\end{eqnarray}
\begin{eqnarray}
T_{k,m}^{k',m'}(k'-1,s,u+1) \, {\cal Q}^{k'}_{k,m-1}(u-s+k') - T_{k,m-1}^{k',m'}(k'-1,s,u+1) \times \qquad \\  %%@
{\cal Q}^{k'}_{k,m}(u-s+k')= T_{k,m}^{k',m'}(k'-1,s+1,u) \, {\cal Q}^{k'}_{k,m-1}(u-s+k'+2), \quad s\leq -m'.
\nonumber
\end{eqnarray}
Taking the second row of ($\overline{BT1}$) (\ref{MTFk'm'}) at $s=-m'+1$ and at $a=k'-1$, the following
equations appear: 
\begin{eqnarray}
T_{k,m}^{k'-1,m'}(a,-m'+1,u) \, {\cal Q}^{k'}_{k,m}(u+a+m'+1) - (-1)^{m'} T_{k,m}^{k',m'}(a,-m'+1,u) \times %%@
\qquad \\  {\cal Q}^{k'-1}_{k,m}(u+a+m'+1)= T_{k,m}^{k',m'}(a+1,-m'+1,u+1) \, {\cal Q}^{k'-1}_{k,m}(u+a+m'-1), %%@
\quad a\geq k',
\nonumber
\end{eqnarray}
\begin{eqnarray}
T_{k,m}^{k',m'}(k'-1,s-1,u+1) \, {\cal Q}^{k'-1}_{k,m}(u-s+k'-1) -T_{k,m}^{k',m'}(k'-1,s,u) \times \qquad \\  %%@
{\cal Q}^{k'-1}_{k,m}(u-s+k'+1)= T_{k,m}^{k'-1,m'}(k'-2,s-1,u) \, {\cal Q}^{k'}_{k,m}(u-s+k'+1), \quad s\leq %%@
-m'.
\nonumber
\end{eqnarray}
Taking the first row of ($\overline{BT2}$) (\ref{MTFk'm'}) at $s=-m'+1$ and at $a=k'-1$, the following
bilinear relations emerge: 
\begin{eqnarray}
T_{k,m}^{k',m'}(k'-1,s,u+1) \, Q^{0,m'-1}(u+s-k') -T_{k,m}^{k',m'-1}(k'-1,s,u+1) \times \qquad \\  %%@
Q^{0,m'}(u+s-k')= T_{k,m}^{k',m'}(k'-1,s-1,u) \,Q^{0,m'-1}(u+s-k'+2), \quad s\leq -m',
\nonumber
\end{eqnarray}
\begin{eqnarray}
T_{k,m}^{k',m'-1}(a+1,-m'+2,u+1) (-1)^{a-k'} Q^{0,m'}(u-a-m')
+ T_{k,m}^{k',m'}(a+1,-m'+1,u) \times \, \\ 
Q^{0,m'-1}(u-a-m'+2)= T_{k,m}^{k',m'}(a,-m'+1,u+1) (-1)^{m'-1} Q^{0,m'-1}(u-a-m'), \, a\geq k'.
\nonumber
\end{eqnarray}

\end{document}